\documentclass[rmp,twocolumn,aps,reprint,nofootinbib]{revtex4}
\usepackage{amssymb}
\usepackage{graphicx}
\usepackage{amsmath}

\def\inbar{\,\vrule height1.5ex width.4pt depth0pt}
\def\IR{\relax{\rm I\kern-.18em R}}
\def\IC{\relax\hbox{$\inbar\kern-.3em{\rm C}$}}

\newcommand{\eq}{\begin{equation}}
\newcommand{\fine}{\end{equation}}

\begin{document}

\title{Investigating macroscopic quantum superpositions
and the quantum-to-classical transition by optical parametric amplification}
\author{Francesco De Martini}
\email{francesco.demartini@uniroma1.it}
\affiliation{Dipartimento di Fisica, Sapienza Universit\`a di Roma, Italy}
\affiliation{Accademia dei Lincei, via della Lungara 10, I-00165 Roma, Italy}
\author{Fabio Sciarrino}
\email{fabio.sciarrino@uniroma1.it}
\affiliation{Dipartimento di Fisica, Sapienza Universit\`a di Roma, I-00185 Roma, Italy}

\begin{abstract}
The present work reports on an extended research endeavor focused on the
theoretical and experimental realization of a macroscopic quantum
superposition (MQS) made up with photons. As it is well known, this
intriguing, fundamental quantum condition is at the core of a famous
argument conceived by Erwin Schroedinger, back in 1935. The main
experimental challenge to the actual realization of \ this object resides
generally on the unavoidable and uncontrolled interactions with the
environment, i.e. the \textquotedblright decoherence\textquotedblright\
leading to the cancellation of any evidence of the quantum features
associated with the macroscopic system. The present scheme is based on a
nonlinear process, the "quantum injected optical parametric amplification",
that maps by a linearized cloning process the quantum coherence of a single
- particle state, i.e. a Micro - qubit, into a Macro - qubit, consisting in
a large number $M$ of photons in quantum superposition. Since the adopted
scheme was found resilient to decoherence, the MQS\ demonstration was
carried out experimentally at room temperature with $M\geq $ $10^{4}$. This
result elicited an extended study on quantum cloning, quantum amplification
and quantum decoherence. The related theory is outlined in the article where
several experiments are reviewed such as the test on the "no-signaling
theorem" and the dynamical interaction of the photon MQS with a
Bose-Einstein condensate. In addition, the consideration of the Micro -
Macro entanglement regime is extended into the Macro - Macro condition. The
MQS\ interference patterns for large $M$ were revealed in the experiment and
the bipartite Micro-Macro entanglement was also demonstrated for a limited
number of generated particles: $M\precsim 12$. At last, the perspectives
opened by this new method are considered in the view of further studies on
quantum foundations and quantum measurement.
\end{abstract}

\maketitle
\tableofcontents

\section{ INTRODUCTION}

\label{sec:intro} Since the golden years of quantum mechanics \cite{Schr35}
the possibility to observe the quantum features of physical systems at the
macroscopic level has been the object of extensive theoretical studies and
recognized as a major conceptual paradigm of physics. However, in general
severe problems stand up to spoil the observation of these features. As it
is well known, the most important one is the unavoidable interaction with
the surrounding environment that determines the loss of any quantum
coherence effect by the corruption of the phase implied by any correlation
of the quantum states~\cite{Zure03}. Such effects are commonly believed to
become increasingly severe with the growing of the size of the system being
studied \cite{raim06rmp}.

In the last several years many experimental attempts have been undertaken to
create superposition of multiparticle quantum states. Different experimental
approaches have been pursued based on atom-photon interacting in a cavity 
\cite{Haro03,raim06rmp}, superconducting quantum circuits \cite{Legg02},
ions \cite{leib03rmp}, micromechanical sytems \cite{mars03prl}, optical
systems \cite{our06sc,our07nat}. In particular in the last few years a
significant advance toward generating superposition states of large objects
using opto-mechanical systems has been achieved \cite%
{grob09nat,roch10nat,teuf11nat}. When dealing with the superposition of
multiparticle quantum states, two are the fundamental issues to be
considered: what is the effective size of the superpositions and how the
state behaves under decoherence phenomena \cite{Legg02}. Several criteria
have been developed to establish the effective size of macroscopic
superpositions in interacting or imperfect scenario, as well as their
applications to real systems \cite{Legg02,dur02prl,kors07pra}. A large
effective size of the state usually conflicts with the robustness of the
quantum superposition under interaction with environment. Moreover the
observation of macroscopic interference phenomena requires to tailor proper
measurement strategies. In particular one faces the problem of achieving a
measurement-precision that enables the observation of quantum effects at
such macro-scales \cite{Kofl07}.

In this paper we discuss how the amplification of quantum states can be
adopted to generate multiphoton superpositions and to investigate the
quantum-to-classical transition. By the present method a quantum
superposition state is first generated in the microscopic (Micro) world of a
single photon particle. Then, such system is mapped into the macroscopic
(Macro)\ realm by generating a quantum superpositions via the well known
"photon stimulation" process of quantum electrodinamics (QED)\ in the regime
of high gain parametric amplification \cite{DeMa98,DeMa98a}. Such approach
is a natural platform for the investigation of the quantum-to-classical
transition, linking quantum and classical matter description. We will review
the properties of the generated states both in the regime of low and high
number of photons. The experimental methods will be oulined and the
corresponding results reported and briefly described. The open question to
devise a method apt to demonstrate experimentally the Micro-Macro
entanglement will be finally addressed.

Let's first consider the regime in which few particles are created by
optical amplification of a single photon in the generic polarization state: $%
\left\vert \phi \right\rangle =\alpha \left\vert H\right\rangle +\beta
\left\vert V\right\rangle $, where $H$ and $V$ stand for the horizontal and
vertical polarization, respectively. This process can be related to several
fundamental tasks of quantum information processing. While classical
information is represented in terms of bits which can be either 0 or 1, the
quantum information theory is rooted on the generation and transformation of
quantum-bits, or qubits, which are two-dimensional quantum systems, each
epitomized by a spins-%
$\frac12$
\cite{Niel00}. A qubit, unlike a classical bit, can exist in a state $%
\left\vert \phi \right\rangle $ that is a superposition of any couple of
orthogonal basis states $\left\{ \left\vert 0\right\rangle ,\left\vert
1\right\rangle \right\} ,$ i.e., $\left\vert \phi \right\rangle =\alpha
\left\vert 0\right\rangle +\beta \left\vert 1\right\rangle $. The fact that
qubits can exist in superposition states gives to quantum information
unusual properties. For instance, a fundamental issue refers to the basic
limitations imposed by quantum mechanics to the set of realizable physical
transformations imposed to the state of any quantum system. The common
denominator of these bounds is that all realizable transformations have to
be represented by completely positive maps which in turn impose a constraint
on the \textquotedblright fidelity\textquotedblright , i.e., the quantum
efficiency, of the quantum measurements. For instance, the fact that an
unknown qubit cannot be precisely determined (or reconstructed) by a
measurement performed on a finite ensemble of identically prepared qubits
implies that this state "cannot be cloned", viz. copied exactly by a general
transformation. In other words, the \textquotedblright
universal\textquotedblright , exact\ cloning map of the form $\left\vert
\phi \right\rangle \rightarrow \left\vert \phi \right\rangle \left\vert \phi
\right\rangle $, or more generally where $N$ are cloned into $M>N$ copies,
is not allowed by the quantum mechanics rules \cite{woot82nat}. Indeed, if
this were possible then one would be able to violate the bound on the
fidelity of estimation and this in turn would trigger most dramatic changes
in the present picture of the physical world. For instance, it would become
possible to exploit the non-local quantum correlations for superluminal
exchange of meaningful information, by then violating the causality
principle \cite{Herb82,DeAn07}. Another map which cannot be performed
exactly on an unknown qubit is the \textquotedblright
spin-flip\textquotedblright , generally dubbed "Universal-NOT"
transformation. This corresponds to the operation $\left\vert \phi
\right\rangle \rightarrow \left\vert \phi ^{\bot }\right\rangle $, where the
state $\left\vert \phi ^{\bot }\right\rangle $ is orthogonal to the original 
$\left\vert \phi \right\rangle $ \cite{Bech99,DeMa02}. The quantum cloning
and the NOT maps are just two amongst a large variety of examples realizing
the effects of the essential limitations imposed by quantum mechanics on
measurements and estimations.

In spite of the fact that these quantum-mechanical transformations on
unknown qubits cannot be performed "exactly", one still may ask what are the
best possible, i.e. \textquotedblright optimal\textquotedblright ,
approximations of these maps within the given structure of quantum theory 
\cite{scar05rmp,cerf06,DeMa05}. In the last few years, it was found possible
to associate an optimal cloning machine with a photon amplification process,
e.g. involving inverted atoms in a laser amplifier or a nonlinear medium in
a "quantum-injected" (QI) "optical parametric amplifier" (OPA) apparatus,
dubbed (QI-OPA) \cite{simo00prl,DeMa00}. In the case of the mode non degenerate QI-OPA, $N$ photons,
prepared identically in an arbitrary quantum state $\left\vert \phi
\right\rangle $ of polarization are injected into the amplifier. By
stimulated emission $M-N$ pairs of photons are created. On the output of the
amplifier generates, in the \textquotedblright cloning
mode\textquotedblright\ are found $M>N$ copies are, or \textquotedblright
clones\textquotedblright\ of the input qubit $\left\vert \phi \right\rangle $
\cite{Lama02,DeMa02,DeMa04}. Correspondingly, the amplifier generates on the
output \textquotedblright anticloning mode", $M-N$ states $\left\vert \phi
^{\bot }\right\rangle $, thus realizing an universal quantum NOT gate \cite%
{DeMa02}.  Moreover the optimal quantum cloning turns out be tightly connected with the no-signaling condition \cite{gisi98pla}.


Let us now address the regime in which a large number of particles is
generated by the amplification process of a single photon in a quantum
superposition state of polarization. Conceptually, the method consists of
transferring the well accessible condition of quantum superposition of a one
photon qubit to a mesoscopic, i.e., multiphoton amplified state $M>1$, here
referred to as a \textquotedblleft mesoscopic qubit\textquotedblright , or
"Macro-qubit". This can be done by injecting in the QI-OPA the one photon
qubit $\alpha |H\rangle +\beta |V\rangle $ \cite{DeMa98,DeMa09,DeMa05a}.
This process is represented in Figure 1 which shows three possible schematic
applications of the method. In virtue of the "information preserving"
(albeit "noisy") property of the amplifier, the generated state is found to
keep the same superposition character and the interfering properties of the
injected qubit. Since the adopted scheme realizes the optimal quantum
cloning machine, able to copy optimally any unknown input qubit into copies
with optimal fidelity, the output state will be necessarily affected by
squeezed-vacuum noise arising from the input vacuum field. We will review
the properties of such Macro states obtained by the quantum injected
amplification process and how they can be exploited to investigate
entanglement in the microscopic-macroscopic (Micro-Macro) regime. Precisely,
there an entangled photon pair is created by a nonlinear optical process;
then one photon of the pair is injected into an optical parametric amplifier
operating for any input polarization state \cite{DeMa08,DeMa09}. Such
transformation establishes a connection between the single photon and the
multiparticle fields. The results of a thorough theoretical analysis
undertaken on this process will be outlined. The results of a series of
related experiments will be reported. We shall show that while a clear
experimental evidence of a MQS interference, in absence of bipartite
Micro-Macro entanglement, has been attained with a fairly large associated
number $M$ of particles, the Micro-Macro entanglement could be consistently
demonstrated, by an attenuation technique only for a small number of
particles: $M$ $\leq 12$. Indeed, as suggested by \cite{seka09prl,Spag10} a
novel "\textit{detection loophole}"\ for large $M$ and the need of very high
measurement resolution impose severe limitations to the detection of quantum
entanglement in the Micro-Macro regime, i.e. of the prerequisite condition
for the full realization of the Schr\H{o}dinger Cat program. In addition, we
shall briefly summarize the potential applications of the QI-OPA technique
in different contexts, such as the realization of a non-locality test,
quantum metrology and quantum sensing.

At last we shall consider a further approach to investigate the
quantum-to-classical transition based on non-linear parametric interactions,
i.e. the one that exploits the process of spontaneous parametric
down-conversion (SPDC) in the high gain regime. In this framework the
investigation of multiphoton states is of fundamental importance, on both
conceptual and practical levels, e.g., for nonlocality tests or for other
quantum information applications . The number of photons generated depends
exponentially on the nonlinear gain $g$ of the parametric process where $g$
can be increased by the adoption of \ high-power pumping lasers and
high-efficiency non-linear crystals. Different experimental approaches to
generate Macro-Macro entangled states and to observe non-local correlations
will be reviewed \cite{Eise04,cami06pra,Vite10,DeMa10fp}. Again the issue of
high-resolution measurements will arise as a fundamental ingredient to
directly observe quantum correlations in the macroscopic regime.

\section{\protect\bigskip OPTICAL PARAMETRIC AMPLIFICATION}

Let us now introduce the non-degenerate optical parametric amplifier which
lies at the core of the present analysis. Consider\ Figure 1 a). Three
different modes of the electromagnetic radiation field - say the signal $%
\widehat{a}_{1}$, the idler $\widehat{a}_{2}$ and the pump $\widehat{a}_{P}$
- are coupled by a non-linear (NL)\ medium, generally a crystal,
characterized by a high third-order tensor expressing the non-linear \
\textquotedblright second-order susceptibility\textquotedblright\ $\chi
^{(2)}$ \cite{Wall95,Yari89,boyd08}. A typical NL\ medium, adopted in the
experiments dealt with in this article, consists of a suitably cut slab of
crystalline \textit{beta barium borate}, commonly dubbed (BBO). Two
\textquotedblright phase matching\textquotedblright\ conditions must be
fulfilled during the coherent three-wave interaction, viz. a scalar one, the
energy conservation, and a vectorial one, the momentum conservation: 
\begin{eqnarray}
\nu _{P} &=&\nu _{1}+\nu _{2}  \label{energyconservation} \\
\overrightarrow{k}_{P} &=&\overrightarrow{k}_{1}+\overrightarrow{k}_{2}
\label{momentumconservation}
\end{eqnarray}%
where the labels $\left\{ P,1,2\right\} $ refer to the \textquotedblright
pump\textquotedblright , signal and idler field modes, respectively. \ The
Hamiltonian of the amplifier under the phase-matched condition (\ref%
{momentumconservation}) can be written in the rotating wave approximation as
follows 
\begin{equation}
\hat{\mathcal{H}}=ik\hbar\left( \hat{a}_{1}^{\dagger }\hat{a}_{2}^{\dagger }%
\hat{a}_{P}+\hat{a}_{1}\hat{a}_{2}\hat{a}_{P}^{\dagger }\right)
\label{hamichi2}
\end{equation}%
The first term of the Hamiltonian (\ref{hamichi2}) describes the physical
process in which a photon is annihilated at frequency $\nu _{P}$ and the
twin photons are generated at frequencies $\nu _{1}$ and $\nu _{2}$. The
second term corresponds to the inverse process. In exact phase-matching
condition the parameter $k$ is proportional to the crystal $\chi ^{(2)}$ and
to the effective crystal length $l_{crist}$ \cite{Yari89,boyd08}.

The Hamiltonian in Eq.(\ref{hamichi2}) describes also the frequency
degenerate case, case in which the frequencies associated with the modes $%
\hat{a}_{1}$ and $\hat{a}_{2}$ are equal but the respective wave-vectors
and/or polarizations are different. The quantum dynamics determined by the
Hamiltonian (\ref{hamichi2}) leads to a rich variety of phenomena, such as
generation of strongly correlated photon pairs by parametric down-conversion 
\cite{shih88prl,Ou88,rari90prl}, quantum-injected optical parametric
amplification \cite{DeMa98}\ phase insensitive amplification \cite{moll67pr}%
, generation of polarization entanglement \cite{kwia95prl}. The unitary
evolution operator associated with $\hat{\mathcal{H}}$ in the interaction
picture is expressed as: 
\begin{equation}
\widehat{U}=\exp \left[ \tau \left( \hat{a}_{1}^{\dagger }\hat{a}%
_{2}^{\dagger }\hat{a}_{P}+\hat{a}_{1}\hat{a}_{2}\hat{a}_{P}^{\dagger
}\right) \right]
\end{equation}%
where $\tau =kt$, $t$ being the interaction time.

The pump\ field $\ \hat{a}_{P}$ is well described by a coherent state (the
"quasi-classical" Glauber's $\alpha -state)$, generally taken as undepleted
because of the small number of converted photons compared with the very
large number of photons, typically larger than $10^{15}$ associated with
each pump pulse. A precise Manley-Rowe theory accounting for the pump
depletion could be possibly adopted, if necessary. Hence, in the generally
adopted \textquotedblright parametric approximation\textquotedblright\ the
pump mode $\hat{a}_{P}$ is replaced with the complex amplitude of the
corresponding coherent state. In that case the interaction Hamiltonian leads
to the two-mode squeezing operator \cite{Wall95}: 
\begin{equation}
\widehat{S}=\exp \left[ \tau \left( \alpha _{P}\hat{a}_{1}^{\dagger }\hat{a}%
_{2}^{\dagger }+\alpha _{P}^{\ast }\hat{a}_{1}\hat{a}_{2}\right) \right]
\label{squeezingoperator}
\end{equation}

The operator $\widehat{S}$ acting on the vacuum state $\left\vert
0\right\rangle _{1}\left\vert 0\right\rangle _{2}$ creates, via the process
of spontaneous parametric down-conversion (SPDC) the "twin beam state" over
the two spatial output modes $\mathbf{k}_{i}$ $(i=1,2)$ with wavelength $%
\lambda _{i}$: 
\begin{equation}
\widehat{S}\left\vert 0\right\rangle _{1}\left\vert 0\right\rangle _{2}=%
\frac{1}{\cosh \tau }\sum_{n=0}^{\infty }\tau ^{n}\left\vert n\right\rangle
_{1}\left\vert n\right\rangle _{2}
\end{equation}%
The average photon numbers $\overline{n}$ on the two modes are related to
the gain $g=\left\vert \tau \right\vert $ as follows: $\overline{n}=\sinh
^{2}g$. Let us provide some numerical estimate by considering a commonly
adopted apparatus. With a BBO crystal, $1mm$ thick, $\lambda _{P}=400nm$ and 
$\lambda _{1}=\lambda _{2}=800nm$, the efficiency of the SPDC process is
very low, typically around $10^{-15}$.

In general, the pump field can be either a continuous or a pulsed beam \cite%
{DeMa05}. Pulsed lasers are used when a high interaction gain and/or an
exact knowledge of the creation time of a photon pair (a \textquotedblright
biphoton\textquotedblright ) are requested. When this is the case
mode-locked laser beams are adopted with a typical pulse duration of
hundreds of femtoseconds. In the SPDC two different types of phase matching
(either I or II) are used depending on the polarization of the three
interacting fields, i.e. on the character of the corresponding
electromagnetic waves in the birefringent non-linear crystal, whether
\textquotedblright \textit{ordinary wave}\textquotedblright\ $(o)$ or
\textquotedblright \textit{extraordinary wave}\textquotedblright\ $(e)$.
Hereafter we will consider only type II phase-matching, in which signal and
idler are respectively $o$ and $e$ polarized. The spatial distribution of
the emitted SPDC radiation consists of two $\mathbf{k}$-vector cones, one
for each type of wave, having common vertices coinciding with the excited
spot on the NL\ crystal slab, considered very thin. We will restrict for
simplicity to the frequency degenerate case only, i.e. $\nu _{1}=\nu
_{2}=\nu _{p}/2$. In the case of type II\ phase matching two different $%
\mathbf{k}$-vector cones are emitted, the $o-$cone and the $e-$cone having
the same vertex, different axes and intersecting along two straight lines.
The two $\mathbf{k}$-vectors, correlated with different polarizations by the
type II parametric interaction, are parallel to these intersection lines and
belong to different cones. The angle between the axes of these polarization
cones can be changed by a convenient tilting of the NL slab with respect to
the direction of the "pump" beam. When this angle is zero the two $\mathbf{k}
$-vectors overlap, each one keeping his own polarization. This condition
corresponds to the collinear interaction we shall consider shortly, below.

\subsection{Non-collinear amplifier}

The interaction Hamiltonian for the type II amplifier in the noncollinear
regime is given by $\hat{\mathcal{H}}_{U}=\imath \hbar \chi (\hat{a}_{1\psi
}^{\dag }\hat{a}_{2\psi _{\bot }}^{\dag }-\hat{a}_{1\psi _{\bot }}^{\dag }%
\hat{a}_{2\psi }^{\dag })+\mathrm{H.c.}$:\ Figure 1-a). Since this system
possesses a complete SU(2) simmetry, the Hamiltonian maintains the same form
for any simultaneous rotation of the Bloch sphere of the polarization basis
for both output modes $\mathbf{k}_{1}$ and $\mathbf{k}_{2}$. Let us now
analyze the features of this device when adopted in the stimulated emission
by a single photon with polarization $|\psi \rangle $, i.e. in the single -
injection QI-OPA regime. The output state of the amplifier reads:%
\begin{equation}
\begin{aligned} \vert \Phi^{1 \psi,0\psi_{\bot}}_{U} \rangle &= \hat{U}_U
\vert 1 \psi \rangle_{1} = \frac{1}{C^{3}} \sum_{n,m=0}^{\infty}
\Gamma^{n+m} (-1)^{m} \sqrt{n+1} \\ &\vert (n+1)\psi, m\psi_{\bot}
\rangle_{1} \otimes \vert m\psi,n\psi_{\bot} \rangle_{2} \end{aligned}
\label{eq:amplified_state_no_losses}
\end{equation}%
where $C=\cosh g$ , $\Gamma =\tanh g$, and $|p\psi ,q\psi _{\bot }\rangle
_{i}$ stands for a Fock state with $p$ photons polarized $\vec{\pi}_{\psi }$
and $q$ photons polarized $\vec{\pi}_{\psi _{\bot }}$ on spatial mode $%
\mathbf{k}_{i}$. Note that the multi-particle states $|\Phi _{U}^{1\psi
}\rangle$, $| \Phi _{U}^{1\psi ^{\bot }}\rangle $ are orthonormal.

Let us note that previous expression involves superpositions of quantum
states with different photon numbers. Clearly the number of photons in the
pump beam would change slightly, however the pumping beam is a coherent
state with large number of photons hence this variation is negligible and
the pumping beam state can be factorized.

\begin{figure}[t]
\centering
\textbf{(a)} \includegraphics[scale=.40, bb= 14 200 560 607]{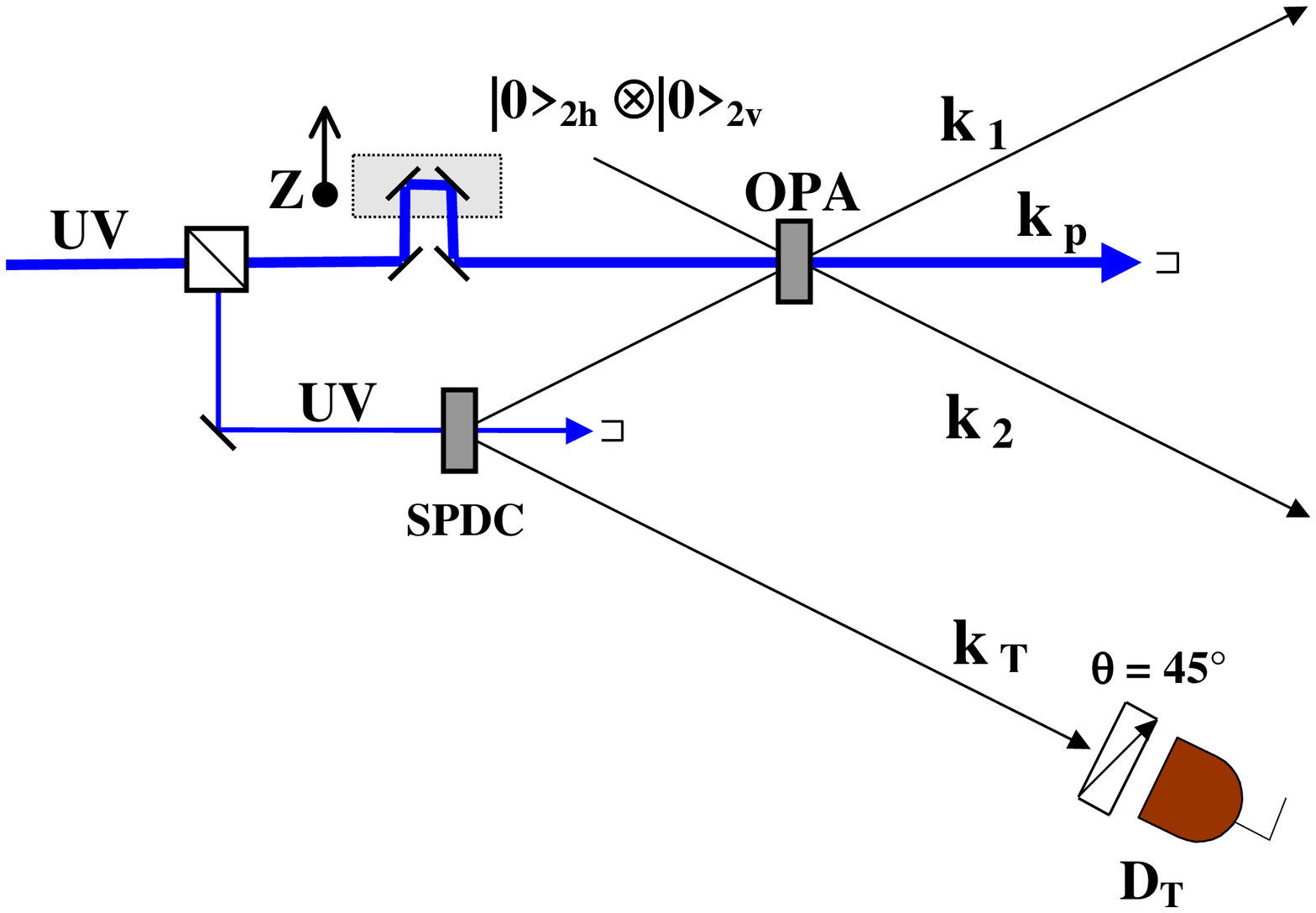} 
\textbf{(b)} \includegraphics[scale=.3]{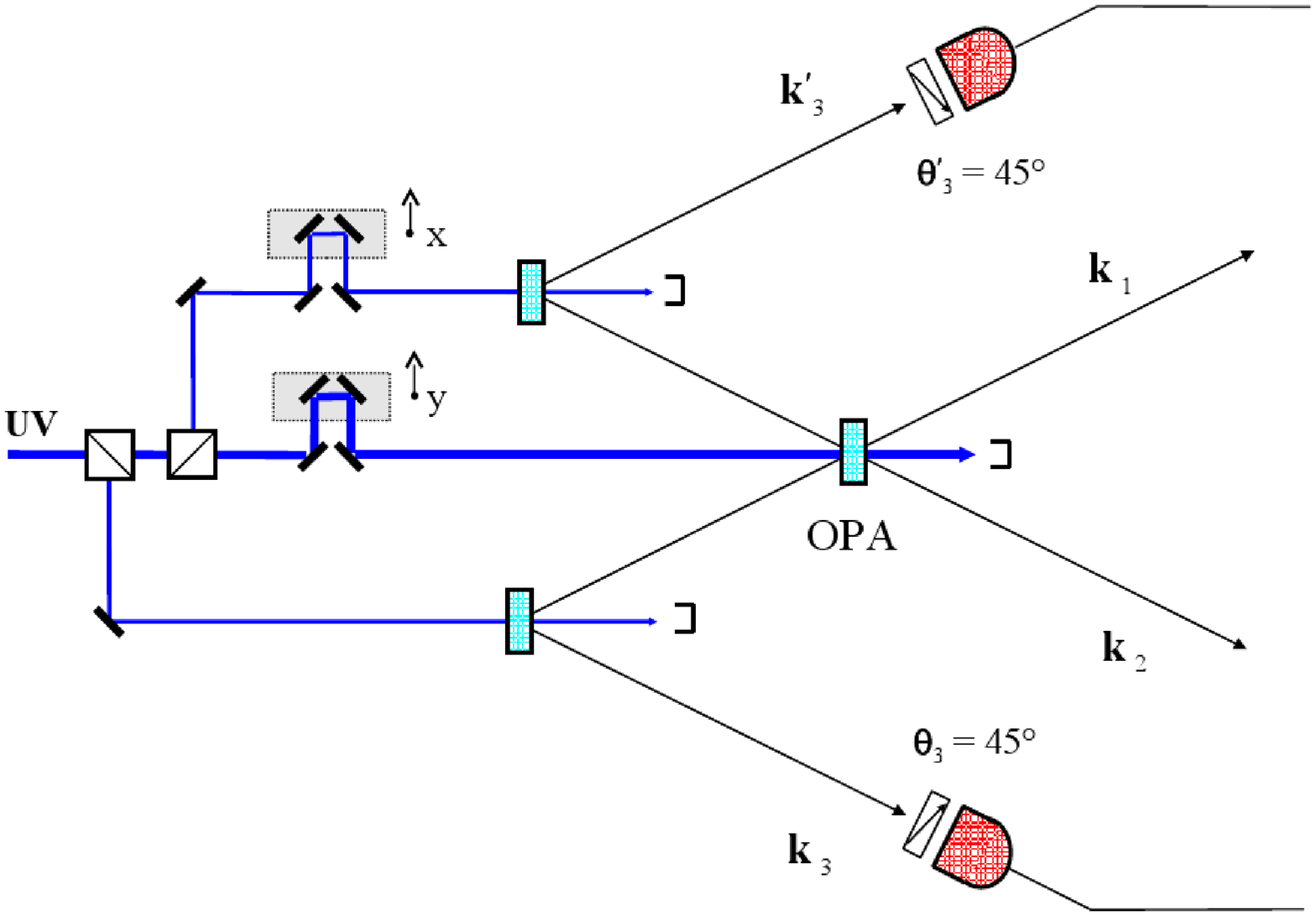} %
\includegraphics[scale=.28]{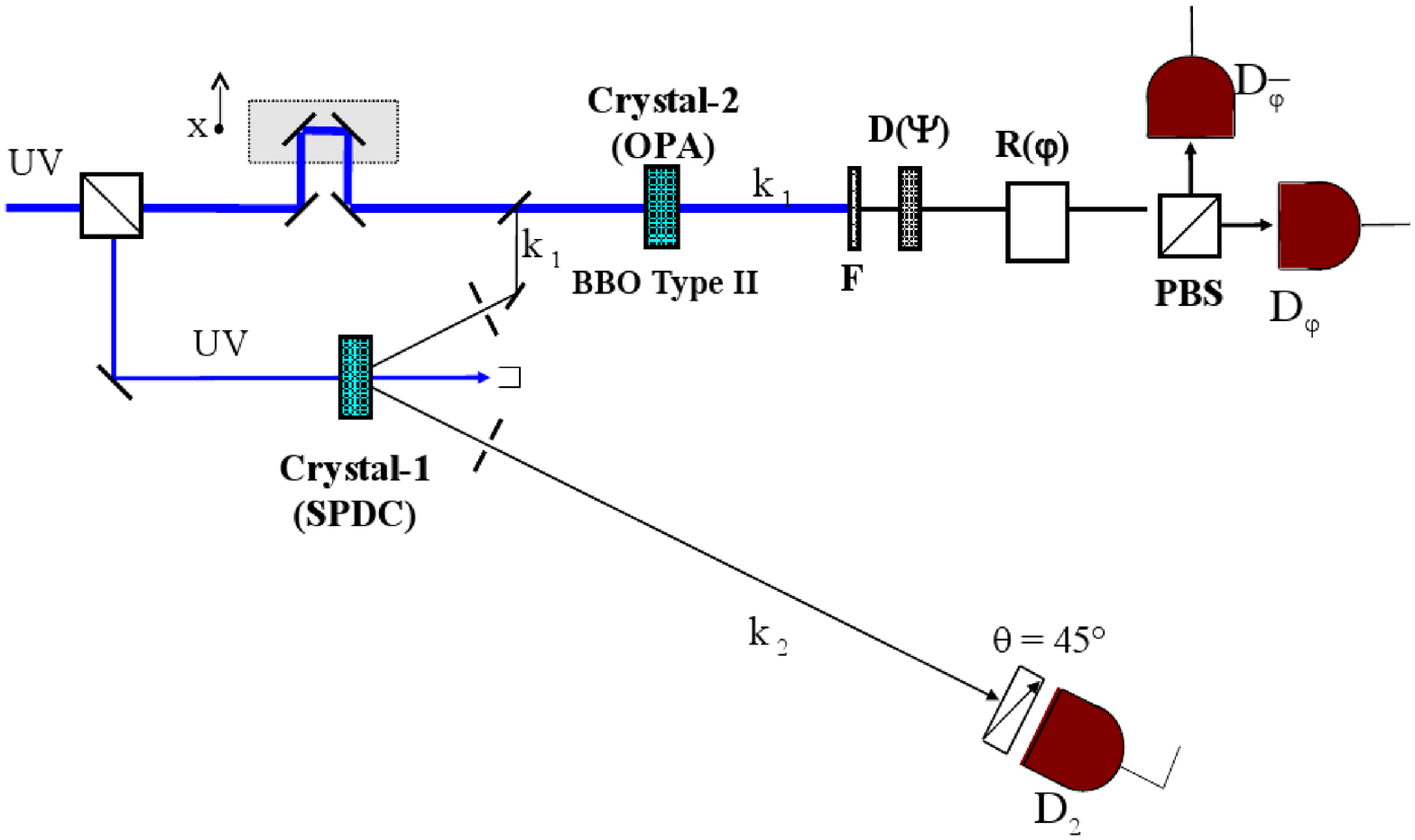} \newline
\textbf{(c)}
\caption{Three different configurations for the amplification of quantum
states. \textbf{(a)} Schematic diagram of the non-collinear quantum injected
optical parametric amplifier. The injection is provided by an external
spontaneous parametric down conversion source of polarization entangled
photon states \protect\cite{DeMa98}.\textbf{(b)} Double-injection of the
optical parametric amplifier \protect\cite{Bovi99}.\textbf{(c)} Collinear
quantum injected optical parametric amplifier \protect\cite{DeMa98a}.}
\label{SchemaInj}
\end{figure}
Let us analyze the output field $\mathbf{k}_{1}$ over the polarization modes 
$\overrightarrow{\pi }_{\pm }$ when the state $\left\vert \varphi
\right\rangle _{1}=2^{-%
{\frac12}%
}(\left\vert H\right\rangle _{1}+e^{i\varphi }\left\vert V\right\rangle
_{1}) $ is injected. The average photon number $M_{\pm }^{i}$ over $\mathbf{k%
}_{i}$ with polarization $\overrightarrow{\pi }_{\pm }$ is found to depend
on the phase $\varphi $ as follows: 
\begin{equation}
M_{\pm }^{1}(\varphi )=\overline{m}+\frac{1}{2}(\overline{m}+1)(1\pm \cos
\varphi )\sinh ^{2}g
\end{equation}%
\begin{equation}
M_{\pm }^{2}(\varphi )=\overline{m}+\frac{1}{2}\overline{m}(1\mp \cos
\varphi )
\end{equation}%
with $\overline{m}=\sinh ^{2}g$. The conditions $\varphi =0$ and $\varphi
=\pi $ correspond to single-photon injection and no-injection on the mode $%
\overrightarrow{\pi }_{+}$, respectively. The average photon number related
to both cases is: $M_{+}^{1}(0)=2\overline{m}+1$ and $M_{+}^{1}(\pi )=%
\overline{m}$.\ The average number of photons emitted over the two
polarizations over $\mathbf{k}_{1}$ is found to be $M=3\overline{m}+1$. The
output state on mode $\mathbf{k}_{1}$ with polarization $\overrightarrow{\pi 
}_{\pm }$ exhibits a sinusoidal fringe pattern of the field intensity
depending on $\varphi $ with a gain-dependent visibility $\mathcal{V}%
_{U}^{th}=\frac{\overline{m}+1}{3\overline{m}+1}$ \cite{DeMa98a}. Note that
for $g\rightarrow \infty $, viz. $M\rightarrow \infty $ the fringe
visibility attains the asymptotic values $\mathcal{V}_{U}^{th}=(\frac{1}{3})$%
. The former considerations are valid for any quantum state injected in the
amplifier $|\phi \rangle $ when analyzed in the polarization basis $\left\{ 
\overrightarrow{\pi }_{\phi },\overrightarrow{\pi }_{\phi \perp }\right\} $.

A more sophisticated extension of the above scheme is the condition of
QI-OPA double-injection represented by Figure 1-b) \ \cite{Bovi99}. Two
separated SPDC sources of polarization entangled photons are adopted to
excite simultaneously over the modes $\mathbf{k}_{1}$ and $\mathbf{k}_{2}$
the QI-OPA amplifier. Meanwhile, the two photons emitted over the external
modes $\mathbf{k}_{3}$ and $\mathbf{k}_{3}^{\prime }$ generate, by a
coincidence circuit, the overall trigger pulse for the experiment when the 
\textit{opposite} polarizations are realized simultaneously. Owing to the NL
dynamics realized by the main NL crystal, the corresponding qubits injected
on the two input QI-OPA\ modes $\mathbf{k}_{1}$ and $\mathbf{k}_{2}$ with
different polarizations give rise to various dynamical processes within the
the QI-OPA amplification. For instance they can lead to an enhanced
interference fringe visibility: $\mathcal{V}_{U}^{th-2}=\frac{2}{3}$.

\subsection{Collinear amplifier}

Let us consider now the results obtained for a collinear optical
configuration in which the two modes $\mathbf{k}_{1}$ and $\mathbf{k}_{2}$
are made to overlap:\ Figure 1-c). The interaction Hamiltonian of this
process is: $\widehat{\mathcal{H}}_{PC}=\imath \hbar \chi \widehat{a}%
_{H}^{\dag }\widehat{a}_{V}^{\dag }+\mathrm{H.c.}$ in the $\left\{ \vec{\pi}%
_{H},\vec{\pi}_{V}\right\} $ polarization basis. The same Hamiltonian is
expressed as $\widehat{\mathcal{H}}_{PC}=\frac{i \hbar \chi }{2}e^{-i \phi
}\left( \widehat{a}_{\phi }^{\dag \,2}-\widehat{a}_{\phi_{\bot }}^{\dag
\,2}\right) +\mathrm{H.c.}$ for any equatiorial basis $\left\{ \vec{\pi}%
_{\phi },\vec{\pi}_{\phi \perp }\right\} $ on the Poincar\'{e} sphere having
as poles the states: $\vec{\pi}_{H},\vec{\pi}_{V}$ . The amplified state for
an injected equatorial qubit $\left\vert \varphi\right\rangle _{1}$ is: 
\begin{equation}
|\Phi _{PC}^{\phi }\rangle =\hat{U}_{PC}|1\varphi,0\varphi _{\bot } \rangle
_{1}=\sum_{i,j=0}^{\infty }\gamma _{ij}|(2i+1)\varphi ,(2j)\varphi _{\bot
}\rangle _{1}  \label{eq:Phi_equat}
\end{equation}%
where $\gamma _{ij}=\frac{1}{C^{2}}\left( e^{-\imath \varphi }\frac{\Gamma }{%
2}\right) ^{i}\left( -e^{-\imath \varphi }\frac{\Gamma }{2}\right) ^{j}\frac{%
\sqrt{(2i+1)!}\,\sqrt{(2j)!}}{i!j!}$, $C=\cosh g$ , $\Gamma =\tanh g$.

The average photon number $M_{\pm }$ over $\mathbf{k}_{1}$ with polarization 
$\overrightarrow{\pi }_{\pm }$ is found to depend on the phase $\varphi $ as
follows: $M_{\pm }(\varphi )=\overline{m}+\frac{1}{2}(2\overline{m}+1)(1\pm
\cos \varphi )$ with $\overline{m}=\sinh ^{2}g$. The average photon number
related to both cases is: $M_{+}(0)=3\overline{m}+1$ and $M_{+}(\pi )=%
\overline{m}$.\ The average number of photons emitted over the two
polarizations over $\mathbf{k}_{1}$ is found to be $M=4\overline{m}+1$.The
sinusoidal fringe pattern of the field intensity has now visibility $%
\mathcal{V}_{PC}^{th}=\frac{2\overline{m}+1}{4\overline{m}+1}$ \cite{DeMa98a}%
. Note that for $g\rightarrow \infty $, the fringe visibility attains the
asymptotic values $\mathcal{V}_{PC}^{th}=(\frac{1}{2})$.

\section{OPTIMAL QUANTUM MACHINES VIA PARAMETRIC AMPLIFICATION}

In the early eighties \cite{Ghir81,woot82nat,Diek82} demonstrated the
impossibility of perfectly copying an unknown arbitrary quantum state. In
other words, a \textquotedblright universal machine\textquotedblright\
mapping $\left\vert \Psi \right\rangle \rightarrow \left\vert \Psi
\right\rangle \left\vert \Psi \right\rangle $ for every $\left\vert \Psi
\right\rangle $ cannot be physically realized. More generally, an exact,
universal cloner of $N$ qubits into $M>N$ qubits cannot exist. Of course
perfect cloning can be provided by a "non-universal" cloning machine, i.e.
one made for one or a restricted class of states. Let us consider the
following scenario: in order \ to copy the quantum state of qubit $C$, we
couple it with another "ancilla" $T$ in the state $|0\rangle $ by adopting a
two qubit logical gate: a Control-NOT (C-NOT). By this approach it is
possible to perfectly copy the state $\left\vert 0\right\rangle _{C}$ or $%
\left\vert 1\right\rangle _{C}$ using the qubit to be copied as control
qubit and an ancilla qubit in the state $\left\vert 0\right\rangle _{T}$\ as
target one \cite{Niel00}. However, starting from any general state $%
\left\vert \phi \right\rangle _{C}=\alpha \left\vert 0\right\rangle
_{C}+\beta \left\vert 1\right\rangle _{C}$ \ the output state generated by
the C-NOT gate is $\alpha \left\vert 0\right\rangle _{C}\left\vert
0\right\rangle _{T}+\beta \left\vert 1\right\rangle _{C}\left\vert
1\right\rangle _{T}$ with $\rho _{C}=\rho _{T}=\left\vert \alpha \right\vert
^{2}\left\vert 0\right\rangle \left\langle 0\right\vert +\left\vert \beta
\right\vert ^{2}\left\vert 1\right\rangle \left\langle 1\right\vert $, which
are clearly different from the initial state $\left\vert \phi \right\rangle
\left\langle \phi \right\vert $. Hence the quantum C-NOT realizes a perfect
cloning machine only for the two input qubit belonging to the set $\left\{
\left\vert 0\right\rangle ,\left\vert 1\right\rangle \right\} .$ Of course
these limitations are effective within the quantum world, i.e. whenever the
quantum superposition character of the state dynamics is a necessary
property of the system, as in an interferometer or, more generally in a
quantum computer.

The no-cloning theorem has also interesting connections with the
impossibility of superluminal communication (generally called "no-signaling
condition") \cite{simo01prl}. That condition will be discussed in details in
Section IV.

We shall see that an approximate, "optimal" solution for cloning, as well
for other quantum processes which are impossible in their "exact" form, is
possible however. By definition, the "optimal" solutions correspond to the
best maps realizable by Nature, i.e., the ones that work just on the
boundaries corresponding to the limitations imposed by the principles of
quantum mechanics.

The concept of optimal cloning process has been first worked out in a
seminal paper by \cite{buze96pra}. A transformation which produces two
copies $(M=2)$ in the same mixed state $\rho _{Cl}$ out of an arbitrary
input qubit $\left\vert \phi \right\rangle $ $(N=1)$ was introduced with a
fidelity equal to 
\begin{equation}
\mathcal{F}_{1\rightarrow 2}\left( \left\vert \phi \right\rangle ,\rho
_{Cl}\right) =\langle \phi |\rho _{Cl}\left\vert \phi \right\rangle =\frac{5%
}{6}  \label{FidClon1to2}
\end{equation}%
This map was demonstrated to be optimal in the sense that it maximizes the
average fidelity between the input state and output qubits in \cite%
{gisi97prl,brus98pra,wern98pra,brus98prl}. More generally, in \cite%
{gisi97prl} a quantum cloning machine has been investigated which transforms 
$N$ identical qubits into $M$ identical copies with an optimal fidelity. In
summary, the universal quantum cloning machine, which transforms $N$
identical qubits $\left\vert \phi \right\rangle $ into $M$ identical copies $%
\rho _{Cl}$, achieves as optimal fidelity: 
\begin{equation}
\mathcal{F}_{N\rightarrow M}(\left\vert \phi \right\rangle ,\rho _{Cl})=%
\frac{N+1+\beta }{N+2}
\end{equation}%
with $\beta \equiv N/M\leq 1\;$\cite{gisi97prl,brus98prl,buze98prl}. It is
useful to compare the previous approach with the process of "state
estimation". Suppose to have $N$ copies of the same quantum state $|\varphi
\rangle $ and to wish to determine all the parameters which characterize $%
|\varphi \rangle $. The optimal estimation procedure leads to a fidelity
between the input state and the estimated one equal to $\mathcal{F}_{est}=%
\frac{N+1}{N+2}$ . As we can see $\mathcal{F}_{N\rightarrow M}(\left\vert
\phi \right\rangle ,\rho _{Cl})$ is larger than the one obtained by the $N$
estimation approach and reduces to that result for $\beta \rightarrow 0$,
i.e. for an infinite number of copies: $M\rightarrow \infty $. The extra
positive term $\beta $ in the above expression accounts for the excess of
quantum information which, originally stored in $N$ states, is optimally
redistributed by entanglement among \ the $M-N$ remaining blank ancilla
qubits\cite{buze96pra}.

In addition to the above results, less \textquotedblright
universal\textquotedblright\ cloning machines have been investigated \cite%
{Buze97,brus98prl}, where the state-dependent cloner is optimal with respect
to a given ensemble of states. As discussed later, this process, generally
referred to as \textquotedblright covariant cloning\textquotedblright ,
operates with a higher fidelity than for the universal cloning since there
is a partial a-priori knowledge of the state (\ref{FidClon1to2}).

The study of optimal quantum cloning is interesting since it implies an
insightful understanding of the critical boundaries existing between
classical and quantum information processing. In the quantum information
perspective, the optimal cloning process may be viewed as providing a
distribution of quantum information over a larger system in the most
efficient way \cite{Ricc05}. More details on the general cloning proces can
be found in the reviews \cite{scar05rmp}, \cite{DeMa05} and \cite{cerf06}.

\subsection{Universal optimal quantum cloning}

Since the first articles on no-cloning theorem it was proposed to exploit
the QED\ stimulated emission process in order to make imperfect copies of
the polarization state of single photons \cite{Milo82,Mand83}. \cite%
{simo00prl,DeMa98,DeMa00} showed that the optimal universal quantum cloning
can indeed be realized by this method. If polarization encoding is adopted,
the \textquotedblright universality\textquotedblright\ of this scheme is
achieved by choosing systems that have appropriate symmetries, i.e., having
a stimulated emission gain $g$ which is polarization independent. This
condition can be achieved by adopting a laser medium or a QI-OPA amplifier
working in the non-collinear configuration. The present Section deals
explicitly with this scheme.

As first scenario we will consider the $1\rightarrow 2$ universal cloning.
Precisely the action of the cloner can be described by the following
covariant transformation \cite{buze96pra}: 
\begin{equation}
\begin{aligned} |\Psi >_{C1}\left| 0 \right\rangle _{C2}\left| 0
\right\rangle _{AC}\Longrightarrow & \sqrt{2/3}|\Psi >_{C1}|\Psi >_{C2}|\Psi
^{\perp }>_{AC} \\&-\sqrt{1/3}|\{\Psi ,\Psi ^{\perp }\}>_{C1,C2}|\Psi >_{AC}
\label{QuantumMachine} \end{aligned}
\end{equation}%
where the first state vector, in the left-hand side of equation (\ref%
{QuantumMachine}), corresponds to the system to be cloned, the second state
vector describes the system on which the information is to be copied
(\textquotedblright blank\textquotedblright\ qubit), represented by the
\textquotedblright \textit{cloning channel}\textquotedblright\ (C), the mode 
$\mathbf{k}_{1}$, while the third state vector represents the cloner
machine. The blank qubit and the cloner are initially in the known state $%
\left\vert 0\right\rangle $. The state $|\{\Psi ,\Psi ^{\perp }\}\rangle $
is the symmetrized state of the two qubit: $2^{-\frac{1}{2}}(|\Psi \rangle
|\Psi ^{\perp }\rangle +|\Psi ^{\perp }\rangle |\Psi \rangle )$.

At the outputs of the cloner $C1$ and $C2$, we find two qubits, the original
and the copy, each one with the following density matrix: 
\begin{equation}
\rho _{C1}=\rho _{C2}=\frac{5}{6}|\Psi ><\Psi |+\frac{1}{6}|\Psi ^{\perp
}><\Psi ^{\perp }|
\end{equation}

The density operators $\rho _{C1}$ and $\rho _{C2}$ describe the best
possible approximation of the perfect universal cloning process. The
fidelity of this transformation does not depend on the state of the input
and is equal to (\ref{FidClon1to2}). The cloner itself after the cloning
transformation is in the state $\rho _{AC}=1/3|\Psi ^{\perp }><\Psi ^{\perp
}|+1/3\times \mathbf{I}$\textbf{\ }, where $\mathbf{I}$ is the unity
operator and is related to the Universal NOT gate, as we shall see later.

Let's now establish a close connection of the above cloning results with the
non-collinear QI-OPA system. The photon injected in the mode $\mathbf{k}_{1}$
has a generic polarization state corresponding to the unknown input qubit $%
\left\vert \Psi \right\rangle $. We shall describe this polarization state
as $\widehat{a}_{\Psi }^{\dagger }\left\vert 0,0\right\rangle
_{1}=\left\vert 1,0\right\rangle _{1}$ where $\left\vert m,n\right\rangle
_{1}$ represents a state with $m$ photons having the polarization $\Psi $,
and $n$ photons with polarization $\Psi ^{\perp }$ on the mode $\mathbf{k}%
_{1}$. Let's assume that the mode $\mathbf{k}_{2}$ is initially in the
vacuum state. The initial polarization state is hence expressed as $%
\left\vert \Psi \right\rangle _{in}=\left\vert 1,0\right\rangle _{1}\otimes
\left\vert 0,0\right\rangle _{2}$ and evolves according to the unitary
operator $\widehat{U}_{U}\equiv \exp \left( -i\widehat{H}_{U}t/\hbar\right) $
(see Section II-A): 
\begin{equation}
\begin{aligned} \widehat{U}_U\left| \Psi \right\rangle _{in}&\simeq \left|
1,0\right\rangle _{1}\otimes \left| 0,0\right\rangle _{2}+\\& g\left(
\sqrt{2}\left| 2,0\right\rangle _{1}\otimes \left| 0,1\right\rangle
_{2}-\left| 1,1\right\rangle _{1}\otimes \left| 1,0\right\rangle _{2}\right)
\label{stimulated} \end{aligned}
\end{equation}%
The linearization procedure implying the above approximation is justified in
the present scenario by the small value of the amplification gain: $g\approx
0.1$ \cite{DeMa02,DeMa04}. The zero order term in Eq.(\ref{stimulated})
corresponds to the process when the input photon in the mode $\mathbf{k}_{1}$
did not interact in the non-linear medium, while the second term describes
the first order amplification process. Here the state $\left\vert
2,0\right\rangle _{1}$ describing two photons of the mode $\mathbf{k}_{1}$
in the polarization state $\Psi $ corresponds to the state $\left\vert \Psi
\Psi \right\rangle $.

\begin{figure}[h!]
\centering \includegraphics[width=0.45\textwidth]{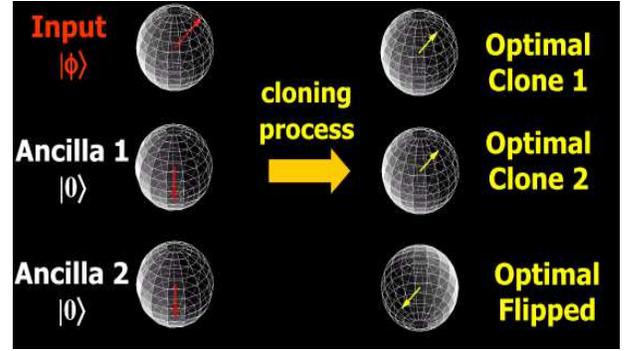}
\caption{Scheme of the optimal cloning process. The input and output qubit
are represented on the Bloch sphere. The vectors associated to the output
states are shrinked compared to the input state $|\protect\phi\rangle$.}
\label{fig:experimental_setup}
\end{figure}

To see that the stimulated emission is indeed responsible for creation of
the cloned qubit, let us compare the state Eq. (\ref{stimulated}) with the
output of the optical parametric amplifier when the vacuum is injected into
the crystal on both input modes $\mathbf{k}_{i}$ ($i=1,2$). In this SPDC\
case the input state is $\left\vert 0\right\rangle _{in}=\left\vert
0,0\right\rangle _{1}\otimes \left\vert 0,0\right\rangle _{2}$, and we
obtain to the same order of approximation as above: 
\begin{equation}
\begin{aligned} \widehat{U}_U\left| 0\right\rangle _{in}&\simeq \left|
0,0\right\rangle _{1}\otimes \left| 0,0\right\rangle _{2}+\\& g\left( \left|
1,0\right\rangle _{1}\otimes \left| 0,1\right\rangle _{2}-\left|
0,1\right\rangle _{1}\otimes \left| 1,0\right\rangle _{2}\right)
\label{spontaneous} \end{aligned}
\end{equation}%
We see that the cloned qubits, described by the state vector $\left\vert
1,0\right\rangle _{1}$ in the right-hand sides of equations (\ref{stimulated}%
) and (\ref{spontaneous}), do appear with different amplitudes,
corresponding to the ratio of the probabilities: $R=2$. It is easy to show
that the fidelity of the output clone is found to be $\frac{2R+1}{2R+2}=%
\frac{5}{6}$ and is optimal.

A more general analysis can be undertaken by extending the isomorphism
discussed above to a larger number of input and output particles $N$ $\ $and 
$M$. In this case it is found that the QI-OPA\ amplification process $%
\widehat{U}_{U}$ in each order of the decomposition into the parameter $g $
corresponds to the $N\longrightarrow M\;$cloning process. Precisely, in this
case $M\geq N$ output particles are detected over the output cloning mode, $%
\mathbf{k}_{1}$. Correspondingly, $M-N$ particles are detected over the
output anticloning mode, $\mathbf{k}_{2}$. The cloning transformation is
realized \textit{a posteriori} in the sense that the output number $M$ of
copies is a random variable that is selected as the result of the
measurement of the photon number in the anticloning beam \cite{simo00prl}.

It appears clear, from the above analysis, that the effect of the input
vacuum field which is necessarily injected in any universal optical
amplifier, is indeed responsible to reduce the fidelity of the quantum
cloning machines at hand. More generally, the vacuum field is in precise
correspondence with, and must be interpreted as, the amount of \ QED\ vacuum
fluctuations that determines the upper bounds to the fidelity determined by
the structure of quantum mechanics.

The universal cloning has been realized by exploiting the stimulated
emission induced by a single photon by \cite{DeMa02,Pell03,DeMa04} as shown
in Figure 3. There a spontaneous parametric down conversion process excited
by the $-\mathbf{k}_{p}$ pump mode created single pairs of photons with
equal wavelengths in entangled singlet states of linear polarization. One
photon of each pair, emitted over $-\mathbf{k}_{1}$, was reflected by a
spherical mirror into the crystal where it provided the $N=1$ single photon
injection into the optical parametric amplifier excited by the pump beam
associated with the backreflected mode $\mathbf{k}_{p}$. Hence the optimal
cloning process was realized along the mode $\mathbf{k}_{1}$. A similar
experiment has been reported by \cite{Lama02} where the single photon
initial qubit was implemented by a highly attenuated coherent state beam.

\begin{figure}[t!!]
\includegraphics[scale=0.45, angle=270]{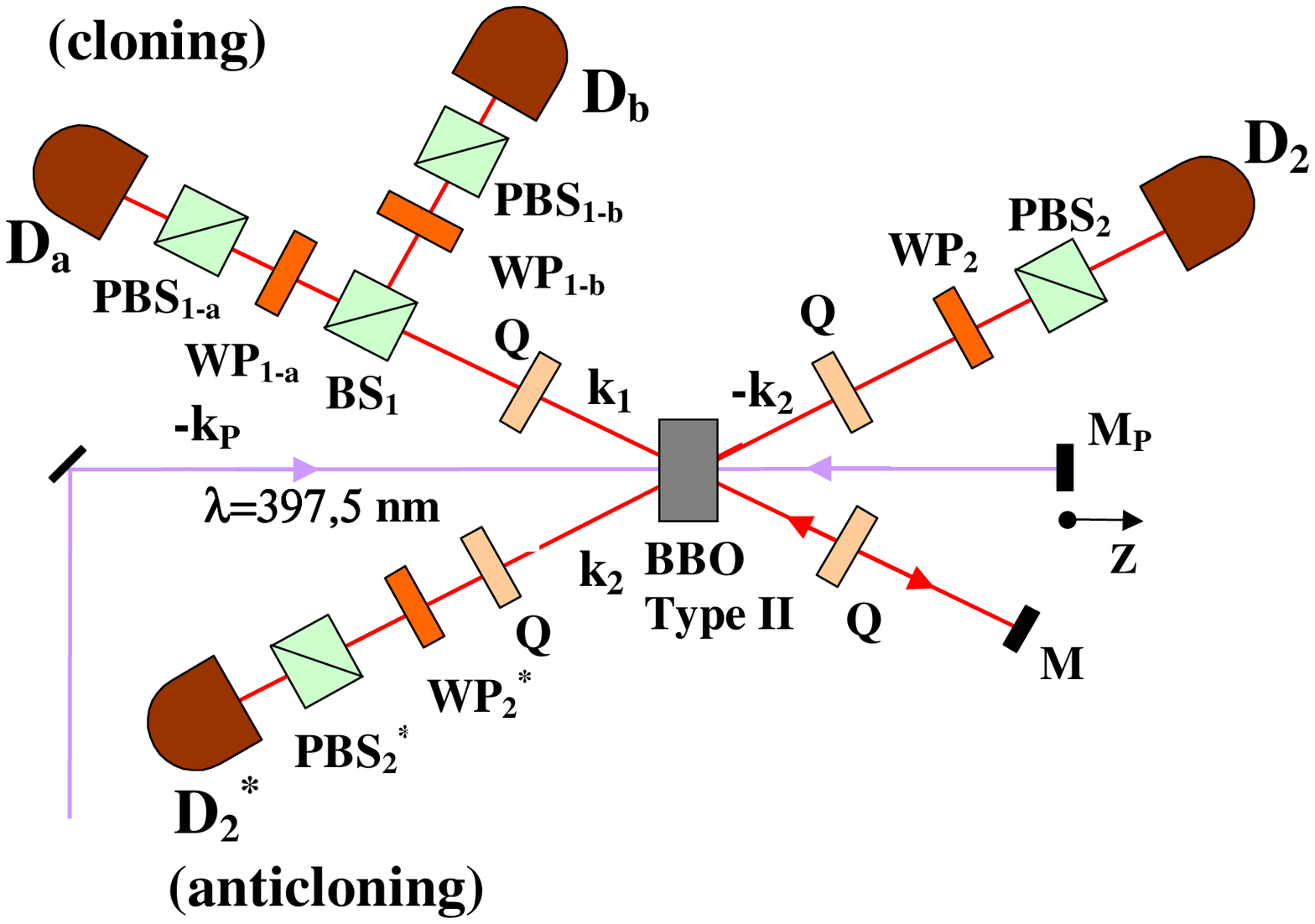}
\caption{Schematic diagram of the \textit{universal optimal cloning machine}
realized on the cloning (C)\ channel (mode $\mathbf{k}_{1}$) of a \textit{%
self-injected} OPA and of the Universal NOT gate realized on the anticloning
(AC)\ channel, $\mathbf{k}_{2}$ \protect\cite{DeMa04}. }
\label{SchemaUNOTCLON}
\end{figure}

\subsection{Universal optimal NOT gate}

The \textquotedblright NOT\ gate\textquotedblright , the transformation that
maps any qubit onto the orthogonal qubit, i.e. onto its antipode on the
Bloch sphere, has been recognized to be impossible according to the
principles of quantum mechanics \cite{Bech99}. In facts if $|\Psi \rangle
=\alpha |0\rangle +\beta |1\rangle $ is a generic qubit, its antipode is
generated by the \textquotedblright time reversal\textquotedblright\
transformation $T\left\vert \Psi \right\rangle $=$\left\vert \Psi ^{\perp
}\right\rangle $ = $\beta ^{\ast }|0\rangle -\alpha ^{\ast }|1\rangle $ such
that $\left\langle \Psi \mid \Psi ^{\perp }\right\rangle =0$. As it is well
known T, being a anti-unitary transformation is not allowed by quantum
mechanics: it may be expressed as: $T=\sigma _{y}K$ \ being $K$ the
transposition or "phase conjugation" map \cite{Niel00}. All this is at
variance with the notion of classical information theory by which the NOT
gate is the simplest operation to be performed "exactly" on any classical
bit. The optimal approximation of the Universal NOT gate (U-NOT) maps $N$
identical input qubits $\left\vert \phi \right\rangle $ into $M$ optimally
flipped ones in the state $\sigma _{out}$ and achieves the fidelity : 
\begin{equation}
\mathcal{F}_{N\rightarrow M}^{\ast }(\left\vert \phi ^{\perp }\right\rangle
,\sigma _{out})=\langle \phi ^{\perp }|\sigma _{out}|\phi ^{\perp }\rangle =%
\frac{N+1}{N+2}  \label{fidUNOT}
\end{equation}%
We note that $\mathcal{F}_{N\rightarrow M}^{\ast }$ depends only on the
number of the input qubits \cite{Gisi99,Buze99,Buze00}. Indeed the fidelity
of the U-NOT gate is exactly the same as the optimal quantum estimation
fidelity \cite{Mass95}. This\ means that the realization of this process is
equivalent to a classical preparation of $M$ identical flipped qubits
following an approximate quantum estimation of $N$ input states. Only this
last operation is affected by noise and in the limit $N\rightarrow \infty $
a perfect estimation of the input state is achieved leading to the
realization of an exact flipping operation.

Let's consider again the expression (\ref{stimulated}) of the output state
of the optical parametric amplifier. The vector $\left\vert 0,1\right\rangle
_{2}$ describes the state of the mode $\mathbf{k}_{2}$ with a single photon
in the polarization state $|\Psi ^{\perp }\rangle $. This state vector
represents the flipped version of the input qubit on mode $\mathbf{k}_{1}$
and then the QI-OPA acts on the output mode $\mathbf{k}_{2}$ as a Universal
NOT-gate \cite{DeMa02}. We see that the flipped qubit described by the state
vector $\left\vert 0,1\right\rangle _{2}$ in the right-hand sides of Eqs.(%
\ref{stimulated}) and (\ref{spontaneous}) do appear with different
amplitudes corresponding to the ratio of probabilities: $R^{\ast }=2:1$.
Note in Eqs.(\ref{stimulated}), (\ref{spontaneous}) that, by calling by $R$
the ratio of the probabilities of detecting $2$ and $1$ photons on mode $%
\mathbf{k}_{1}$\ only, we obtain: $R=R^{\ast }$. In other words, the \textit{%
same value} of signal to noise ratio affects both cloning and U-NOT
processes realized simultaneously on the two different output modes: $%
\mathbf{k}_{1}$ and $\mathbf{k}_{2}$. The corresponding values of the U-NOT
fidelity reads $\mathcal{F}^{\ast }=2/3$ and is equal to the optimal one
allowed by quantum mechanics \cite{DeMa02}.

A remarkable and somewhat intriguing aspect of the present process is that
both processes of quantum cloning and the U-NOT gate are realized
contextually by the \textit{same} physical apparatus, by the same unitary
transformation and correspondingly by the same quantum logic network \cite%
{DeMa04}.

The relation between the cloning and the NOT operations have been latter
discussed according to the conservation laws alone \cite{vane06}. It was
suggested that the close link existing between the limitations on cloning
and NOT operations could express an as yet unexplored natural law. The
result discussed above are general and hold both in classical and
quantum-mechanical worlds, for both optimal and suboptimal operations, and
for bosons as well as fermions.

\subsection{Optimal machines by symmetrization}

Optimal quantum cloning machines, although working probabilistically, have been demonstrated experimentally by a symmetrization technique \cite%
{ricc04prl,scia04pra,scia04pla,irvi04prl}.
This approach to the probabilistic implementation of the $N$ to $M$ cloning process has been first proposed by Werner \cite{wern98pra}. It is based on the action of a projective operation on the symmetric subspace of the $N$ input qubits and $M-N$ blank ancillas. This transformation assures the uniform distribution of the initial information into the overall system and guarantees that all output qubits are indistinguishable. 
To achieve the projection over the symmetric subspace we exploit the bosonic
nature of photons, viz. the exchange symmetry of their overall wavefunction. In particular we use a two-photon Hong-Ou-Mandel coalescence effect \cite{hong87prl}. In this process, two photons impinging simultaneously on a
beamsplitter from two different input modes have an enhanced probability of
emerging along the same output mode (that is, "\textit{coalescing}"), as
long as they are indistinguishable. If the wo photons are made
distinguishable, e.g by different encoding of their polarization or of any
other degree of freedom, the coalescence effect vanishes. Now, if one of the
two photons involved in the process is in a known input state to be cloned
and the other is in a random one, the coalescence effect will enhance the
probability that the two photons emerge from the beamsplitter with the same
quantum state. In other words, the symmetrization enhances the probability
of a successful cloning detected at the output of the beamsplitter.

The universal optimal quantum cloning based on the symmetrization technique
was first demonstrated for polarization encoded qubit  \cite%
{ricc04prl,scia04pra,scia04pla,irvi04prl} and latter reported for orbital angular momentum-encoded qubits \cite%
{naga09npho}. Finally \cite{naga10prl} reported the experimental realization
of the optimal quantum cloning of four-dimensional quantum states, or
ququarts, encoded in the (polarization + orbital angular momentum) degrees
of freedom of photons \cite{marr11}.

\subsection{Phase-covariant optimal quantum cloning}

In addition to the impossibility of universally cloning unknown qubits,
there exists the impossibility of cloning subsets of qubits containing non
orthogonal states. This no-go theorem has been adopted to provide the
security of cryptographic protocols as $BB84$ \cite{Gisi02}. Recently state
dependent, non universal, optimal cloning machines have been investigated
where the cloner is optimal with respect to a given ensemble \cite{Brub00}.
This partial a-priori knowledge of the state allows to reach a larger
fidelity than for the universal cloning.

The simplest and most relevant case is represented by the cloning covariant
under the Abelian group $U(1)$ of phase rotations, the so called
"phase-covariant cloning". There, the information is encoded in the phase $%
\phi _{i}$ of the input qubit belonging to the equatorial plane $i$ of the
corresponding Bloch sphere. In this context the general state to be cloned
may be expressed as: $\left\vert \phi _{i}\right\rangle =(\left\vert \psi
_{i}\right\rangle +e^{i\phi _{i}}\left\vert \psi _{i}^{\perp }\right\rangle
) $ and $\left\{ \left\vert \psi _{i}\right\rangle ,\left\vert \psi
_{i}^{\perp }\right\rangle \right\} $ is a convenient orthonormal basis \cite%
{Brub00}. The values of the optimal fidelities $\mathcal{F}%
_{cov}^{N\rightarrow M}$ for the phase-covariant cloning machine have been
found in \cite{DAri03}. Restricting the analysis to a single input qubit to
be cloned $N=1$ into $M>1$ copies, the cloning fidelity is found: $\mathcal{F%
}_{cov}^{1\rightarrow M}=\frac{1}{2}\left( 1+\frac{M+1}{2M}\right) $ for $M$
assuming odd values, or $\mathcal{F}_{cov}^{1\rightarrow M}=\frac{1}{2}%
\left( 1+\frac{\sqrt{M\left( M+2\right) }}{2M}\right) \;$for $M$ even$.$ In\
particular we have $\mathcal{F}_{cov}^{1\rightarrow 2}=0.854$ and $\mathcal{F%
}_{cov}^{1\rightarrow 3}=0.833$\ to be compared with the corresponding
figures valid for universal cloning: $\mathcal{F}_{univ}^{1\rightarrow
2}=0.833$ and $\mathcal{F}_{univ}^{1\rightarrow 3}=0.778$. It is worth to
enlighten the deep connection linking the phase-covariant cloning and the
estimation of an equatorial qubit, that is, with the problem to finding the
optimal strategy to estimate the value of the phase $\phi $ \cite{derk98prl}%
. In general for $M\rightarrow \infty ,$ $\mathcal{F}_{cov}^{N\rightarrow
M}\rightarrow \mathcal{F}_{phase}^{N}.$ In particular we have $\mathcal{F}%
_{cov}^{1\rightarrow M}=\mathcal{F}_{phase}^{1}+\frac{1}{4M}$ with $\mathcal{%
F}_{phase}^{1}=3/4.$

\begin{figure}[h]
\includegraphics[scale=.3]{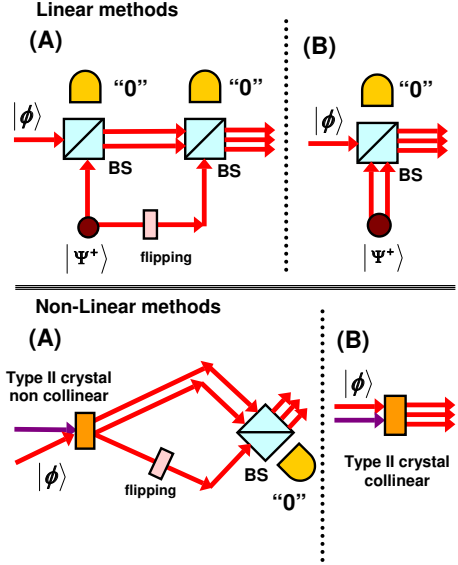}
\caption{Linear methods: (\textbf{a}) schematic diagram of the linear optics
multi qubit symmetrization apparatus realized by a chain of interconnected
Hong-Ou-Mandel interferometer; (\textbf{b}) symmetrization of the input
photon and the ancilla polarization entangled pairs. Non-linear methods: (%
\textbf{a}) UQCM by optical parametric amplification, flipping by a couple
of waveplates and projection over the symmetric subspace; (\textbf{b})
collinear optical parametric amplification \protect\cite{Scia07}.}
\label{figCov}
\end{figure}

We shall briefly review different schemes which can be realized through the
methods of quantum optics outlined above \cite{Scia07}. Let us restrict our
attention to the $1\rightarrow 3$ phase-covariant quantum cloning machine as
the corresponding scheme can be easily extended to general case $%
1\rightarrow M$ for odd values of $M$. The phase-covariant cloner can be
realized by adopting a quantum-injected optical parametric amplifier
(QI-OPA)\ working in a collinear configuration: Figure 1 c) \cite{DeMa98a}.
In this case the interaction Hamiltonian $\hat{\mathcal{H}}_{PC}=i\chi \hbar
\left( \widehat{a}_{H}^{\dagger }\widehat{a}_{V}^{\dagger }\right) +h.c.$
acts on a single output spatial mode $\mathbf{k}$. A fundamental physical
property of $\hat{\mathcal{H}}_{PC}$ consists of its rotational invariance
under $U(1)$ transformations, that is, under any arbitrary rotation around
the $z$-axis. Let us consider an injected single photon with polarization
state $\left\vert \phi \right\rangle _{in}=2^{-1/2}(\left\vert
H\right\rangle +e^{i\phi }\left\vert V\right\rangle )=\left\vert
1,0\right\rangle _{k}$where $\left\vert m,n\right\rangle _{k}$ represents a
product state with $m$ photons of the mode $\mathbf{k}$ with polarization $%
\phi $, and $n$ photons with polarization $\phi ^{\perp }$. The first
contribution to the amplified state, $\sqrt{6}\left\vert 3,0\right\rangle
_{k}-\sqrt{2}e^{i2\phi }\left\vert 1,2\right\rangle _{k}$ is identical to
the output state obtained fron a $1\rightarrow 3$ phase-covariant cloning.
Indeed the fidelity is found to be the optimal one $F^{1\rightarrow 3}=\frac{%
5}{6}$ \cite{Scia07}. Notice the effect of the input vacuum field over the single \textbf{k} mode with polarization $\phi ^{\perp }$ coupled to the phase-covariant optical amplifier, this vacuum contribution is indeed responsible to reduce the fidelity of the quantum cloning machine. 

Interestingly, the same overall state evolution can also be obtained at the
output of a non-collinear QI-OPA together with a Pauli $\sigma _{Y}$
operation and the projection of the three output photons over the symmetric
subspace Fig.\ref{figCov}(\textbf{a}). This scheme was experimentally
realized by the following method: the flipping operation on the output mode $%
\mathbf{k}_{AC}$ was realized by means of two waveplates, while the physical
implementation of the symmetrization projector on the three photons-states
was carried out by linearly superimposing the modes $\mathbf{k}_{C}$ and $%
\mathbf{k}_{AC}$ on a beamsplitter $BS$ and then by selecting the case in
which the three photons emerged from $BS$ all on the same output mode $%
\mathbf{k}_{PC}$\cite{Scia05}.


\section{PARAMETRIC AMPLIFICATION AND NO-SIGNALING THEOREM}

In the present paragraph we review the connections between the cloning
process and the special theory of relativity according to which any signal
carrying information cannot travel at a speed larger than the velocity of
light in vacuum:$\ c$. Even if quantum physics has marked nonlocal features
due to the existence of entanglement, it has been found that a "no-signaling
theorem" exists according to which one cannot exploit quantum entanglement
between two space-like separated parties for "faster-than-light"
communication \cite{maud02}. Several attempts to break this "peaceful
coexistence" have been proposed, the most renowned one by Nick Herbert in
1981 by its FLASH machine (First Laser-Amplified Superluminal Hookup) \cite%
{Herb82}.The publication of this proposal, based on a cloner machine applied
to an entangled state of two spacelike distant particles $A$ and $B$, was
followed by a lively debate that eventually stimulated the formulation of
the "no-cloning theorem"\cite{woot82nat}.

\begin{figure}[h]
\includegraphics[scale=.52]{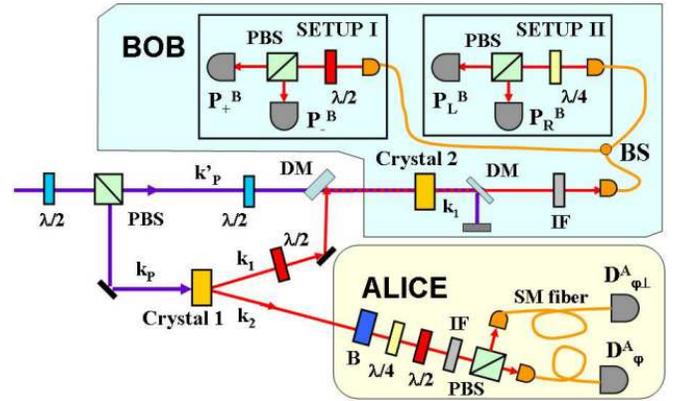}
\caption{Configuration of the quantum injected optical parametric amplifier.
The SPDC quantum injector (crystal 1) is provided by a type II generator of
polarization entangled photon couples \protect\cite{DeAn07}.}
\label{FLASH}
\end{figure}

The setup proposed by Herbert is reported in Fig.\ref{FLASH}. If one
observer, $Bob$ by measuring the particle $B$ could distinguish between
different state mixtures that have been prepared by the distant observer $%
Alice$ by measuring the particle $A$, then quantum non-locality could be
used for signaling. Precisely, let's consider the following: Alice and Bob
share two polarization entangled photons $A$ and $B$ generated by a common
source. Alice detects her photon polarization with the detectors $D_{\varphi
}^{A}$ and $D_{\varphi \bot }^{A}$ either in the basis$\{\overrightarrow{\pi 
}_{\pm }=2^{-1/2}(\overrightarrow{\pi }_{V}\pm \overrightarrow{\pi }_{H})\}$
or $\{\overrightarrow{\pi }_{R}=2^{-1/2}(\overrightarrow{\pi }_{H}+i%
\overrightarrow{\pi }_{V}),\overrightarrow{\pi }_{L}=2^{-1/2}(%
\overrightarrow{\pi }_{V}-i\overrightarrow{\pi }_{H})\}$, where $%
\overrightarrow{\pi }_{H}$ and $\overrightarrow{\pi }_{V}$ are,
respectively, linear horizontal and vertical polarization. If Bob could
guess with a probability larger than 
$\frac12$
the basis chosen by Alice, superluminal signaling would be established. It
was recognized that this is impossible if the experiment involves two single
particles. However Herbert thought that Bob could make a \textquotedblright
new kind\textquotedblright\ of measurement involving the amplification of
the received signal $B$\ trough a \textquotedblright nonselective laser gain
tube\textquotedblright , viz, a universal (polarization independent)
amplifier. The amplified photon beam is "split" by an optical "beam
splitter" (BS), so Bob can perform on one of the two output channels of BS a
measurement on half of the cloned particles by an apparatus tuned on the
basis $\left\{ \overrightarrow{\pi }_{\pm }\right\} $ who records the signal 
$I_{\pm }^{B}$. Simultaneously, he could record on the other BS output
channel the signals $I_{R/L}^{B}$by an apparatus tuned on the basis $\left\{ 
\overrightarrow{\pi }_{R},\overrightarrow{\pi }_{L}\right\} $. In that way
he could guess the right preparation basis carried out by Alice\ on particle
A.

In order to test the Herbert's scheme a careful theoretical and experimental
analysis of the output field was carried out with emphasis on fluctuations
and correlations \cite{DeAn07}. Precisely, the Herbert's scheme was
reproduced by the optical parametric amplification of a single photon of an
entangled pair into an output field involving $5\times 10^{3}$ photons. Fig.%
\ref{FLASH}. Unexpected and peculiar field correlations amongst the cloned
particles preventing any violation of the no-signaling conditions have been
found. Precisely, it was found that the limitations implied by a complete
quantum cloning theory are not restricted to the bounds on the cloning
fidelity but also largely affect the high-order correlations existing among
the different clones. In fact, in spite of a reduced fidelity, noisy but
separable, i.e., noncorrelated, copies would lead to a perfect state
estimation for $g\rightarrow \infty $ and hence to a real possibility of
superluminal communication. However, surprisingly enough the particles
produced by any optimal cloning machine are highly interconnected and the
high-order correlations are actually responsible for preventing any
possibility of faster than light communication \cite%
{demk05pra,bae06prl,DeAn07}. Recently
\cite{Zhan11} investigated the wavepacket propagation of a single photon
and showed experimentally in a conclusive way that the single photon speed is 
limited by $c$.
\begin{widetext}

\begin{figure}[t!!]
\centering
\includegraphics[width=1\textwidth]{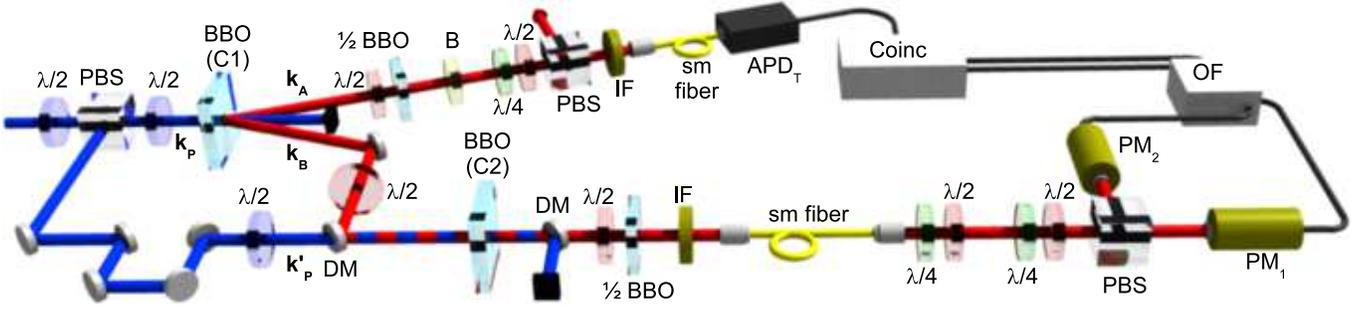}
\caption{Scheme of the experimental setup. The main ultraviolet laser beam
provides the excitation field beam at $\lambda_{P} = 397.5 nm$. A type II
Beta Barium Borate crystal (crystal 1: C1) generates pair of photons with $\lambda = 795 nm$.
The photon belonging to $\mathbf{k}_{B}$, together with the pump laser beam
$\mathbf{k}_{p}'$, is fed into an high gain optical parametric amplifier consisting
of a crystal 2 (C2), cut for collinear type-II phase matching. Measurement apparatus:
the field is analyzed by two photomultipliers (PM$_{1}$ and PM$_{2}$) and then discriminated through an
O-Filter device (OF), whose action is described in the text. For more details refer to \cite{DeMa08}.}
\label{fig:experimental_setup}
\end{figure}

\end{widetext}

\section{EXPERIMENTAL\ MACROSCOPIC QUANTUM SUPERPOSITION BY MULTIPLE\
CLONING OF SINGLE PHOTON STATES}

\subsection{Generation and detection of multi-particle quantum superpositions%
}

The present Section accounts for the optical parametric amplification of a
single photon in the high gain regime to experimentally investigate how the
information initially contained in its polarization state is distributed
over a large number of particles. In particular we analyze how the coherence
properties of the input state are transferred to the mesoscopic output field.

Let's consider the scenario in which a single-particle qubit $|\psi \rangle
_{B}=\alpha \left\vert \phi \right\rangle _{B}+\beta \left\vert \phi ^{\bot
}\right\rangle _{B}$ , with $\left\vert \alpha \right\vert ^{2}+\left\vert
\beta \right\vert ^{2}=1$, injected in a three-wave optical parametric
amplifier \cite{Yari89} is transformed by the unitary QI-OPA operation into
a corresponding macroscopic quantum superposition (MQS): 
\begin{equation}
\left\vert \Phi \right\rangle _{B}=\alpha \left\vert \Phi ^{\phi
}\right\rangle _{B}+\beta \left\vert \Phi ^{\phi \perp }\right\rangle _{B}
\end{equation}%
The multi-particle states, or macrostates, whose detailed expression
reported in Eq.\ref{eq:Phi_equat}, bear peculiar properties that deserve
some comments. The macrostates, $\left\vert \Phi ^{\phi }\right\rangle _{B}$%
, $\left\vert \Phi ^{\phi \perp }\right\rangle _{B}$ are orthonormal and
exhibit observables bearing macroscopically distinct average values.
Precisely, the average number of photons associated with the polarization
mode $\overrightarrow{\pi }_{\phi }$ is: $\overline{m}=\sinh ^{2}g$ for $%
\left\vert \Phi ^{\phi \perp }\right\rangle _{B}$, and $(3\overline{m}+1)$
for $\left\vert \Phi ^{\phi }\right\rangle _{B}$. For the $\pi -$mode $%
\overrightarrow{\pi }_{\phi \perp }$ , orthogonal to $\overrightarrow{\pi }%
_{\phi }$, these values are interchanged among the two states. On the other
hand, as shown by \cite{DeMa98}, by changing the representation basis from $%
\left\{ \overrightarrow{\pi }_{\phi },\overrightarrow{\pi }_{\phi \perp
}\right\} \;$to $\left\{ \overrightarrow{\pi }_{H},\overrightarrow{\pi }%
_{V}\right\} $, the same macro-states, $\left\vert \Phi ^{\phi
}\right\rangle _{B}$ or $\left\vert \Phi ^{\phi \perp }\right\rangle _{B}$
are found to be again quantum superpositions of two orthonormal states $%
\left\vert \Phi ^{H}\right\rangle _{B}$, $\left\vert \Phi ^{V}\right\rangle
_{B}$, but differing by a single quantum. This unexpected and quite peculiar
combination, i.e. a large difference of a measured observable when the
states are expressed in one basis and a small Hilbert-Schmidt distance of
the same states when expressed in another basis turned out to be a
fundamental property that renders the coherence properties of the system
robust toward the coupling with environment. This will be discussed later in
Section VII. .

Let us first briefly review the adopted optical system adopted making
reference to the experimental layout shown by \ the sketchy Figure 1 (c), or
by the equivalent, more detailed Figure 6\cite{cami06las,Naga07,DeMa08}. An
entangled pair of two photons in the \ singlet state $\left\vert \psi
^{-}\right\rangle _{A,B}$=$2^{-{\frac{1}{2}}}\left( \left\vert
H\right\rangle _{A}\left\vert V\right\rangle _{B}-\left\vert V\right\rangle
_{A}\left\vert H\right\rangle _{B}\right) $, was produced through
spontaneous parametric down-conversion (SPDC) by the BBO\ crystal 1 (C1)
pumped by a (weak) pulsed ultraviolet pump beam: Fig.\ref%
{fig:experimental_setup}. There the labels $A,B$ refer to particles
associated respectively with the two output spatial modes $\mathbf{k}_{A}$%
and $\mathbf{k}_{B}$ of the SPDC generated by C1. In the experiment the
three spatial modes involved in the injected parametric interaction were
carefully selected adopting single mode fibers. Consequently, in virtue of
Equations (1, 2) a three-wave, collinear "phase-matching" condition was
realized leading to a lossless amplification process.

\bigskip The single photon qubit on mode $\mathbf{k}_{A}$ of Fig. 6 (i.e. $%
\mathbf{k}_{2}$ in Fig. 1(c)) was sent to a "polarizing beam splitter" (PBS)
whose output modes were coupled to two single-photon detectors. In a first
experiment, these two detectors were simply connected as to merely identify
the emission of \ $A$ (and of $B$) without any effective polarization
measurement on $A$. In other words, the two detectors acted as the single
detector unit $D_{2}$ of the simplified Figure 1 c) by supplying a single
electronic "trigger signal" for the overall experiment in correspondence
with the emission of any entangled couple $A,B~$\ emitted by C1. In virtue
of this "trigger signal" the overall measurement of the MQS $\left\vert \Phi
\right\rangle _{B}$ could be considered "heralded by", i.e. measured in
coincidence with any photon $A$ measured on mode $\mathbf{k}_{A}$.
Accordingly, no bipartite Micro-Macro entanglement was detectetable\cite%
{DeMa08,DeMa98a}.


\ The single particle qubit $|\psi \rangle _{B}=2^{-\frac{1}{2}}(|H\rangle
_{B}-|V\rangle _{B})$ associated with particle $B$, prepared in the
superposition state of polarization state was then injected, together with a
very intense laser pulsed pump beam, into the main optical parametric
amplifier consisting of a second BBO\ crystal 2 (C2). The crystal C2 was
oriented for "\textit{collinear operation}", i.e., emitting pairs of
amplified photons over the same output spatial mode supporting two
orthogonal polarization modes, respectively horizontal and vertical. The
high - gain parametric amplification provided by the crystal C2 transformed
the single - particle qubit: $|\psi \rangle _{B}$ into the corresponding
multi-particle MQS: $\left\vert \Phi \right\rangle _{B}=2^{-\frac{1}{2}%
}(\left\vert \Phi ^{\phi }\right\rangle _{B}-\left\vert \Phi ^{\phi \perp
}\right\rangle _{B})$. This MQS, composed of $M\approx 5\times 10^{4}$
photons, was analyzed\ by a "polarization analyzer" [A($\varphi $)] coupled
through a PBS\ to two high-gain photomultipliers (PM)\ with detection
efficiency $\simeq $ $5\%$. The device A($\varphi $) composed by a optical
rotator and of a birefringent plate analyzed the MQS $\left\vert \Phi
\right\rangle _{B}\ $in a rotating base characterized by a single $\varphi -$%
phase. More details are given in:\cite{DeMa98a}.

\begin{figure}[h]
\includegraphics[scale=.27]{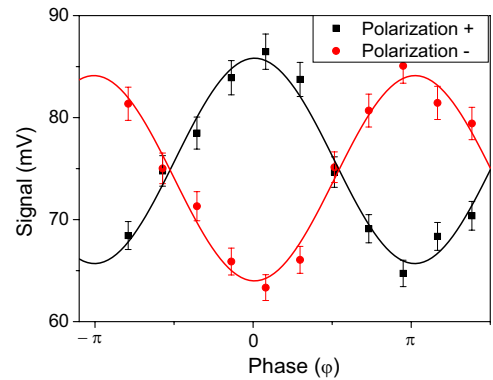}
\caption{Average signal versus the phase of the input qubit \protect\cite%
{Naga07}.}
\label{SchemaInj}
\end{figure}
The sinusoidal behaviour shown by the detected experimental
interference-fringe pattern reported in Fig.\ref{SchemaInj} as function of \ 
$\varphi $, offers a visual realization of the original 1935 pictorial
argument by Erwin Schr\H{o}dinger \cite{Schr35}. These interference-fringe
patterns show how the coherent quantum superposition properties of the input
state can be transferred to the mesoscopic output, involving up a very large
number of photon particles. There the minima and the maxima of the patterns,
e.g. shown by Fig.7, could be attributed to the dead/alive conditions of the
celebrated Schr\H{o}dinger feline, i.e of the Macro-system. Similar MQS\
interference fringe - patterns arising in different experiments have been
obtained: e.g. as shown by Figure 9, below and by Figure 4 of \cite{DeAn07}.
As already mentioned, all MQS results were obtained at "room temperature"
thus defying the "phase-disrupting" decoherence process that generally
affects this kind of experiment. However we shall see that the observation
of other sophisticated quantum effects such as the entanglement correlations
within Micro-Macro systems requires not only a system well protected against
environmental decoherence, but also a sufficient measurement resolution. We
shall see also that, in spite of the reported successful evidence of the
MQS\ realization, the measurement of bipartite Micro-Macro entanglement with
very large $M\ $meets severe experimental problems, owing to a newly
discovered multiparticle "efficiency loophole"\cite{seka09prl}. However, by
a different quantum tomographic test within a "deliberate attenuation"
experiment carried out in a non-collinear configuration, the achievement of
bipartite Micro-Macro entanglement was demonstrated for a limited number of
\ QI-OPA\ generated particles: $M\leq 12$ \cite{DeMa05a} The last
experiment, demonstrating the achievement of bipartite Micro-Macro
entanglement as well as of the MQS condition, will be discussed below, in
Sections VI and IX.

\bigskip 

\begin{figure}[h]
\includegraphics[scale=.8]{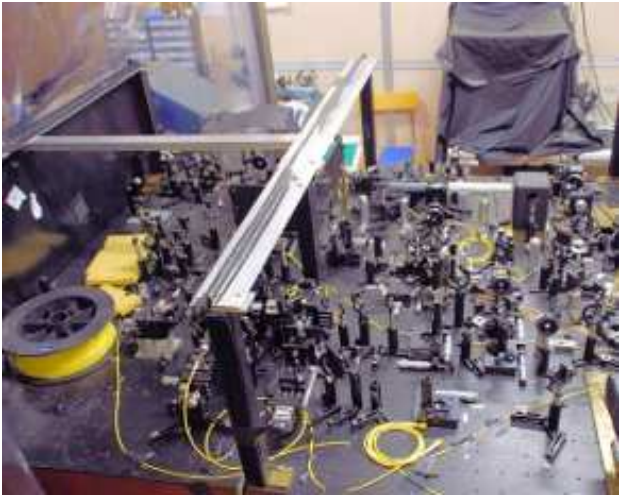}
\caption{Picture of the experimental apparata adopted for the amplification
of entangled pairs of photons (Quantum Optics Group, Dipartimento di Fisica,
Sapienza Universita di Roma). }
\label{SchemaInj}
\end{figure}

\section{MICRO-MACRO SYSTEM: HOW TO DEMONSTRATE ENTANGLEMENT}

Owing to the close similarity existing between the QI-OPA\ scheme shown in
Figure 1-c) and the well known EPR\ scheme it may be argued that the
non-local separability between the single photon qubit $\left\vert \psi
^{-}\right\rangle _{A,B}$=$2^{-{\frac{1}{2}}}\left( \left\vert
H\right\rangle _{A}\left\vert V\right\rangle _{B}-\left\vert V\right\rangle
_{A}\left\vert H\right\rangle _{B}\right) \ $emitted over the output mode $\ 
\mathbf{k}_{B}$ and the Macro-state emitted over $\mathbf{k}_{A}~$could be
demonstrated experimentally\ \cite{cami06las}. Formally, this endeavor would
consist of the demonstration of the existence of the entangled state
connecting the two Micro-Macro systems $A$ and $B$:

\begin{equation}
|\Psi ^{-}\rangle _{AB}=\frac{1}{\sqrt{2}}(|\phi \rangle _{A}|\Phi ^{\phi
_{\bot }}\rangle _{B}-|\phi _{\bot }\rangle _{A}|\Phi ^{\phi }\rangle _{B})
\end{equation}%
There the output macro state is expressed by $|\Phi ^{\phi }\rangle =\hat{U}%
_{PC}|\phi \rangle $, where $|\phi \rangle $ labels the injection of
single-photon state is (\ref{eq:amplified_state_no_losses}).

Such a demonstration would consist of a complete physical achievement of the
1935 Schr\"{o}dinger Cat program. In the following subsections different
theoretical and experimental approaches will be briefly discussed. In
particular, an ambitious attempt in this direction was undertaken with a
high gain QI-OPA method generating a Macro state consisting of nearly $%
M=10^{4}$ photons. The experimental layout was similar to the one described
in Section V but adopted a different, sophisticated processing of the
signals generated by the particle detectors \cite{DeMa08}. However, it was
soon realized that a conclusive test of Micro-Macro entanglement for a very
large number of particles could only be achieved successfully by adoption of
linear photomultipliers featuring a very large detection efficiency $\eta
\lesssim 1$, a condition not made available by the present technology\cite%
{seka09prl}. This is but the effect of a new form of the well known "\textit{%
detection loophole}" which\ affects in general all nonlocality tests and is
found to worsen for an increasing number $M\ $of detected particles.
However, as said, in spite of all these problems a conclusive experimental
demonstration of Micro-Macro entanglement has been realized by a quantum \
tomographic method for a limited number of MQS\ particles: $M\lesssim 12$,
as we shall immediately see in the next paragraph.

\subsubsection{Extracted two photon density matrices}

A feasible approach for the analysis of multiphoton fields is based on the
deliberate attenuation of the analyzed system up to the single photon level 
\cite{Eise04}. In this way, standard single-photon techniques and criteria
can be used to investigate the properties of the field. The verification of
the bi-partite entanglement in the high loss regime is an evidence of the
presence of entanglement before attenuation, on the premise that no
entanglement can be generated by any "local operations", including lossy
attenuation. The attenuation method has been applied to the Micro-Macro
system, realizing by a quantum tomographic method the experimental proof of
the presence of entanglement between the single photon state of mode $%
\mathbf{k}_{A}$ and the multiphoton state with $M\lesssim 12$ of mode $%
\mathbf{k}_{B}$ generated through the process of parametric amplification in
an universal cloning configuration:\ Figure 6. The theory of this experiment
will be considered once again, in more details in Section IX, below\cite%
{DeMa05a,cami06pra}. Unfortunately, such approach could only be applied for
a very limited number $M$, since in practice unavoidable experimental
imperfections quickly wash out any evidence of entanglement \cite{Spag10}.

\subsubsection{Pseudo-spin operators}

Let us now address a different criteria to verify the bipartite entanglement
between the modes $\mathbf{k}_{A}$ and $\mathbf{k}_{B}$. We adopt the
standard Pauli operators for the single photon polarization state belonging
to mode $\mathbf{k}_{A}$. We introduce a formalism useful to associate the
amplified multi-particle field on mode $\mathbf{k}_{B}$ to a Macro-qubit.
Through the amplification process the spin operators $\hat{\sigma}_{i}$ of
the single photon evolve into the $"$macro-spin' operators $\hat{\Sigma}_{i}$%
\ for the many particle system $\hat{\Sigma}_{i}=\hat{U}\hat{\sigma}_{i}\hat{%
U}^{\dagger }=\left\vert \Phi ^{\psi i}\right\rangle \left\langle \Phi
^{\psi i}\right\vert -\left\vert \Phi ^{\psi i\perp }\right\rangle
\left\langle \Phi ^{\psi i\perp }\right\vert .$ The operators $\left\{ \hat{%
\Sigma}_{i}\right\} $ satisfy the same commutation rules of the single
particle $\frac{1}{2}-$spin $\left[ \hat{\Sigma}_{i},\hat{\Sigma}_{j}\right]
=\varepsilon _{ijk}2i\hat{\Sigma}_{k}$ where $\varepsilon _{ijk}$ is the
Levi-Civita tensor density. Hence the generic state $\alpha \left\vert \Phi
^{H}\right\rangle _{B}+\beta \left\vert \Phi ^{V}\right\rangle _{B}$ can be
handled as a qubit in the Hilbert space $H_{B}$ spanned by $\left\{
\left\vert \Phi ^{H}\right\rangle _{B},\left\vert \Phi ^{V}\right\rangle
_{B}\right\} $. To test whether the output state is entangled, one should
measure the correlation between the single photon spin operator $\hat{\sigma}%
_{i}^{A}$ on the mode $\mathbf{k}_{A}$ and the macro-spin operator $\widehat{%
\Sigma }_{i}^{B}$, on the mode $\mathbf{k}_{B}$. We then adopt the criteria
for two qubit bipartite systems based on the spin-correlation. We define the
visibility $V_{i}=\left\vert \left\langle \widehat{\Sigma }_{i}^{B}\otimes 
\widehat{\sigma }_{i}^{A}\right\rangle \right\vert $ a parameter which
quantifies the correlation between the systems $A$ and $B$. The value $%
V_{i}=1$ corresponds to perfect anti-correlation, while $V_{i}=0$ expresses
the absence of correlation. The following upper bound criterion for a
separable state holds \cite{Eise04}: $S=\sum_{i}V_{i}\leq 1$. In order to
measure the expectation value of $\widehat{\Sigma }_{i}^{B}$, a
discrimination among the pair of states $\left\{ \left\vert \Phi ^{\psi
i}\right\rangle ,\left\vert \Phi ^{\psi i\perp }\right\rangle \right\} $ for
the three different polarization bases $1,2,3$ is required.\ Consider the
two macro-states $\left\vert \Phi ^{+}\right\rangle $ and $\left\vert \Phi
^{-}\right\rangle $. A perfect discrimination can be achieved by identifying
whether the number of photons over the $\mathbf{k}_{B}$ mode with
polarization $\overrightarrow{\pi }_{+}$ is even or odd. As already said,
this requires the detection of the mesoscopic field by a
photon-number-resolving detectors operating with an overall quantum
efficiency $\eta $ $\approx $ 1, a device not yet made available by the
present technology. We face here the problem of detecting correlations by
performing a coarse-grained measurement process.

\begin{figure}[h]
\includegraphics[scale=.3]{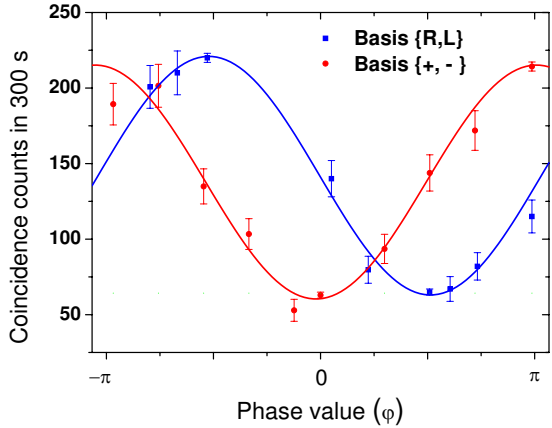}
\caption{Coincidence counts $[L_B; D_A]$ versus the phase of the injected
qubit for the diagonal (circle data) and circular polarization basis(square
data) \protect\cite{DeMa08}. }
\label{SchemaInj}
\end{figure}

\subsubsection{Correlation measurements via orthogonality filter}

In order to implement a measurement with high-discrimination, a new method
has been adopted, viz. the O-Filter (OF) based strategy. This method is
based on a probabilistic discrimination of the macro-states $|\Phi ^{\phi
}\rangle $ and $|\Phi ^{\phi _{\bot }}\rangle $, which exploits the
macroscopic features present in their photon-number distributions \cite%
{Naga07}. Such measurement is implemented by an intensity measurement
carried out by multiphoton linear detectors in the $\{\vec{\pi}_{i},\vec{\pi}%
_{i}^{\bot }\}$ basis, followed by an electronic processing of the recorded
signal. If $n_{\pi }-m_{\pi _{\bot }}>k$, the (+1) outcome is assigned to
the event, if $m_{\pi _{\bot }}-n_{\pi }>k$ the (-1) outcome is assigned to
the event. If $|n_{\pi }-m_{\pi _{\bot }}|<k$, an inconclusive outcome (0)
is assigned to the event.

Experimentally the photon is detected on mode $\mathbf{k}_A$ adopting single
photon detectors and the multiphoton field of mode $\mathbf{k}_B$ with
photomultipliers and O-Filter. The experimental fringe patterns shown in Fig.%
\ref{SchemaInj} were obtained by adopting the common analysis basis $\left\{ 
\overrightarrow{\pi }_{R},\overrightarrow{\pi }_{L}\right\} \;$with a
filtering probability $\simeq 10^{-4}$. In this case the average visibility
has been found $V_{2}=(54.0\pm 0.7)\%$. A similar oscillation pattern has
been obtained in the basis $\left\{ \overrightarrow{\pi }_{+},%
\overrightarrow{\pi }_{-}\right\} $ leading to: $V_{3}=\left( 55\pm 1\right)
\%$. Since always is $V_{1}>0$, the experimental result $S\;$= $%
V_{2}+V_{3}\; $= $(109.0\pm 1.2)\%$ should imply the violation of the
separability criteria introduced above. However a carefull analysis of the
implications of discarding part of the data via the OF measurement should be
addressed.

The state after losses is no more a macro-qubit living in a two dimensional
Hilbert space, but in general it is represented by a density matrix $\hat{%
\rho}_{\eta }^{\phi }$. A detailed discussion on the properties of the
macrostates after losses in both the Fock-space and the phase-space is
reported in \cite{DeMa09,Spag09,Spag10}. Very generally, the probabilistic
detection method described above can be adopted to infer the active
generation \textsl{before} losses of a Macro-state $|\Phi ^{\phi }\rangle ,$
or $|\Phi ^{\phi _{\bot }}\rangle $, by exploiting the information encoded
in the unbalancement of the number of particles present in the state \textsl{%
after} losses $\hat{\rho}_{\eta }^{\phi }$. Hence the adopted entanglement
criterion allowed to infer the presence of bipartite\ Micro-Macro
entanglement present before losses, under a specific assumption \cite{DeMa08}%
.\ This point was discussed extensively by \cite{seka10pra,seka09prl} who
showed that any loss of data allows the formulation of a kind of "detection
loophole" that impairs the success of the entanglement demonstration. Let's
remind that it has been known for at least four decades that a general "%
\textit{detection loophole}" exists in the refutation of Local Realistic
Theories, and is the source of skepticism about the definitiveness of all
experiments dealing with single particle Bell Inequality violations. In
facts, as claimed repeatedly by John S. Bell itself , because of the absence
of an experimental confirmation of the\ "fair sampling" assumption or of a
plausible equivalent one, all experimental tests of the Bell's Inequality
may today interpreted in large areas of the scientific community as merely
\textquotedblleft good indications\textquotedblright\ of the real existence
of quantum non-locality \cite{bell87,gree86,maud02}. However, it is also
well known that the detection loophole can be closed for single particle
Bell's inequality experiments by the adoption of detectors with efficiency
as large as $\eta \geq 85\%$ \cite{eber93pra}. Quite unfortunately our
results show that an even larger value of $\eta $ is required to demonstrate
Micro-Macro entanglement in multiparticle systems. A thorough analysis of
the Micro-Macro entanglement was carried out by \cite{Spag10} demonstrating
that a priori knowledge of the system that generates the Micro-Macro pair is
necessary to exclude a class of separable states that can reproduce the
obtained experimental results. In conclusion, the genuine unbiased
demonstration of bipartite Micro-Macro entanglement, i.e. in absence of any
a priori assumption, is still an open experimental challenge when a very
large number $M\ $of particles are involved.


\subsubsection{Effects of coarse-grained measurement}

Recently \cite{raei11qph} analyzed the effects of coarse-graining in photon
number measurements on the observability of Micro-Macro entanglement that is
created by greatly amplifying one photon from an entangled pair. They
compared the results obtained for a unitary quantum cloner, which generates
Micro-Macro entanglement, and for a measure-and-prepare cloner, which
produces a separable Micro-Macro state. Their approach demonstrates that the
distance between the probability distributions of results for the two
cloners approaches zero for a fixed moderate amount of coarse-graining. Once
again, this proves that the detection of Micro-Macro entanglement becomes
progressively harder as the system's size increases \cite{raei11qph}.

\begin{figure}[b!]
\centering
\includegraphics[width=0.45\textwidth]{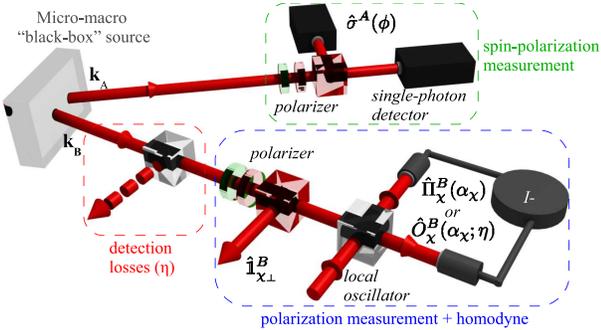}
\caption{(Color online) Hybrid non-locality and entanglement test on an
optical microscopic-macroscopic state generated by a ``black-box''. The
single-photon mode $\mathbf{k}_{A}$ is measured by a polarization detection
apparatus, while the multi-photon mode $\mathbf{k}_{B}$ undergoes both
polarization and homodyne measurements.}
\label{fig:schemaConceptual}
\end{figure}

As alternative approach to demonstrate the Micro-Macro entanglement, \cite%
{raei11bqph} proposed a scheme where a photon is first cloned using
stimulated parametric down conversion, making many optimal copies, and then
the cloning transformation is inverted, regenerating the original photon
while destroying the copies. Focusing on the case where the initial photon
is entangled with another photon, \cite{raei11bqph} studied the conditions
under which entanglement can be proven in the final state. This proposed
experiment would provide a clear demonstration that quantum information is
preserved in phase-covariant quantum cloning but again one photon should be
lost between the cloning trasformation and the following inversion process.
The experimental reversion of the optimal quantum cloning and flipping
processes has been reported by \cite{scia06pra}. There, the combination of
linear and nonlinear optical methods was exploited to implement a scheme
that, after the cloning transformation, restores the original input qubit in
one of the output channels, by using local measurements, classical
communication, and feedforward. This nonlocal method demonstrated how the
information on the input qubit can be restored after the cloning process.

\subsubsection{\protect\bigskip Hybrid criteria}

Very recently \cite{Spag11} analyzed a hybrid approach to the experimental
assessment of the genuine quantum features of a general system consisting of
microscopic and macroscopic parts: Fig.\ref{fig:schemaConceptual}. They inferred the presence of
entanglement by combining dichotomic measurements on a bidimensional system
and a phase-space inference through the Wigner distribution associated with
the macroscopic component of the state. As a benchmark, the method was
adopted to investigate the feasibility of the entanglement demonstration in
a bipartite-entangled state composed of a single-photon and a multiphoton
field. This analysis shows that, under ideal conditions, maximal violation
of a Clauser-Horne-Shimony-Holt inequality is achievable regardless of the
number of photons $M\ $in the macroscopic part of the state. The problems
arising in the detection of entanglement when losses and detection
inefficiency are included can be overcome by the use of a hybrid
entanglement witness that allows efficient correction for losses in the
few-photon regime. This analysis elicits further interest in the
identification of suitable test in the high-loss and large-photon-number
region and paves the way to an experimentally feasible demonstration of \
the properties of entanglement affecting a quite interesting class of states
lying at the very border between the quantum and the classical domains.

\section{RESILIENCE TO DECOHERENCE OF THE AMPLIFIED MULTIPARTICLE STATE}

In this Section we discuss the resilience to decoherence of the quantum
states generated by optical parametric amplification of a single-photon
qubit. The basic tools of this investigation are provided by two coherence
criteria expressed by \cite{DeMa09,DeMa09a}. There, the Bures distance \cite%
{Bure69,Hubn92,Jozs94}: 
\begin{equation}
\mathcal{D}\left( \hat{\rho},\hat{\sigma}\right) =\sqrt{1-\sqrt{\mathcal{F}(%
\hat{\rho},\hat{\sigma})}},
\end{equation}%
where $\mathcal{F}$ is a quantum "fidelity", has been adopted as a measure
to quantify: (\textbf{I}) the $"$\emph{distinguishability"} between two
quantum states $\left\{ |\phi _{1}\rangle ,|\phi _{2}\rangle \right\} $ and (%
\textbf{II)} the \emph{degree of coherence}, i.e. superposition visibility
of their macroscopic quantum superpositions (MQS) $|\phi ^{+}\rangle
=2^{-1/2}\left( |\phi _{1}\rangle \pm |\phi _{2}\rangle \right) $. These
criteria were chosen according to the following considerations: \textbf{(I)}
The distinguishability i.e. the degree of orthogonality, represents the
maximum discrimination among two quantum states achievable within a
measurement. \textbf{(II)} The visibility, between the superpositions $|\phi
^{+}\rangle $ and $|\phi ^{-}\rangle $ depends exclusively on the relative
phase of the component states:$\ |\phi _{1}\rangle $ and $|\phi _{2}\rangle $%
. Consider two orthogonal superpositions $|\phi ^{\pm }\rangle $: $\mathcal{D%
}\left( |\phi ^{+}\rangle ,|\phi ^{-}\rangle \right) =1$. In presence of
decoherence the state evolves according to a phase-damping channel $\mathcal{%
E}$, the relative phase between $\ |\phi _{1}\rangle $ and $|\phi
_{2}\rangle $ progressively randomizes and the superpositions $|\phi
^{+}\rangle $ and $|\phi ^{-}\rangle $ approach an identical fully mixed
state leading to: $\mathcal{D}\left( \mathcal{E}(|\phi ^{+}\rangle ),%
\mathcal{E}(|\phi ^{-}\rangle \right) )=0$. The physical interpretation of $%
\mathcal{D}\left( \mathcal{E}(|\phi ^{+}\rangle ),\mathcal{E}(|\phi
^{-}\rangle \right) )$ as visibility is legitimate insofar as the component
states of the corresponding superposition, $|\phi _{1}\rangle $ and $|\phi
_{2}\rangle $ may be defined, at least approximately, as \textquotedblleft 
\textit{pointer states}\textquotedblright\ or \textquotedblleft \textit{%
einselected states}\textquotedblright\ \cite{Zure03}. Within the set of the
eigenstates characterizing the system under investigation, the pointer
states are defined as the ones less affected by the external noise and that
are highly resilient to decoherence.

Let's now compare the resilience properties of the different classes of
quantum states after the propagation over a lossy channel $\mathcal{E}$.
This one is modelled by a linear beam-splitter (BS) with transmittivity $T$
and reflectivity $R=1-T$ acting on a state $\widehat{\rho }$ associated with
a single BS input mode. Let's first analyze the behaviour of the coherent
states and their superpositions. The investigation on the Glauber's states
leading to the $\alpha -MQS^{\prime }s$ case \cite{Schl91}:$|\Phi _{\alpha
\pm }\rangle $=$\mathcal{N}^{-1/2}\left( |\alpha \rangle \pm |-\alpha
\rangle \right) $ in terms of the \textquotedblleft pointer states" $|\pm
\alpha \rangle \ $leads to the closed form result:\ $\mathcal{D}(\mathcal{E}%
(|\Phi _{\alpha +}\rangle ),\mathcal{E}(|\Phi _{\alpha -}\rangle ))=\sqrt{1-%
\sqrt{1-e^{-4R|\alpha |^{2}}}}$ . This one is plotted in Fig.\ref%
{fig:resilience} (dashed line) as function of \ the average number of lost
photons: $x\equiv R\langle n\rangle $. Note that the value of $\mathcal{D}(%
\mathcal{E}(|\Phi _{\alpha +}\rangle ),\mathcal{E}(|\Phi _{\alpha -}\rangle
))$ drops from 1 to 0.095 upon loss of only one photon: $x=1$. In other
words, any superposition of $\alpha -states.$ $|\Phi _{\alpha \pm }\rangle $=%
$\mathcal{N}^{-1/2}\left( |\alpha \rangle \pm |-\alpha \rangle \right) $\
exhibits a fast decrease in its coherence, i.e. of its "visibility" and
"distinguishability", while the related components $|\pm \alpha \rangle $,
i.e. the \textquotedblleft pointer states\textquotedblright\ \cite{Zure03},
remain distinguishable until all photons of the state are depleted by the BS.

Let us now analyze the behaviour of the amplified multiphoton states by a
QI-OPA\ apparatus described in the previous Sections. An EPR pair $|\psi
^{-}\rangle =2^{-1/2}\left( |H\rangle _{A}|V\rangle _{B}-|V\rangle
_{A}|H\rangle _{B}\right) $ is generated in a first non-linear crystal:\
Figure 6. By analyzing and measuring the polarization of the photon
associated with the mode $\mathbf{k}_{A}$, the photon on mode $\mathbf{k}%
_{B} $ is prepared in the polarization qubit: $|\psi \rangle _{B}=\cos
(\theta /2)|H\rangle _{B}+e^{\imath \phi }\sin (\theta /2)|V\rangle _{1}$.
Then, the single photon is injected in the amplifier simultaneously with the
strong UV pump beam $\mathbf{k^{\prime }}_{P}$. Let us analyze the two
configurations of the apparatus leading, as said, to two different regimes
of quantum cloning:\ the "phase covariant" and the "universal".

\subsection{Phase-covariant optimal quantum cloning machine}

We evaluated numerically the \textit{distinguishability} of $\left\{ |\Phi
_{PC}^{+,-}\rangle \right\} $ through the distance $\mathcal{D}(\mathcal{E}%
(|\Phi _{PC}^{+}\rangle ),\mathcal{E}(|\Phi _{PC}^{-}\rangle ))$: Fig.\ref%
{fig:resilience}. Consider the MQS of the macrostates $|\Phi
_{PC}^{+}\rangle $, $|\Phi _{PC}^{-}\rangle $: $\left\vert \Phi
_{PC}^{R/L}\right\rangle =\frac{\mathcal{N}_{\pm }}{\sqrt{2}}\left( |\Phi
_{PC}^{+}\rangle \pm i|\Phi _{PC}^{-}\rangle \right) $. Due to the linearity
of the amplification process and in virtue of the phase-covariance of the
process \cite{DeMa09,DeMa09a}:

\begin{equation}
\mathcal{D}(|\Phi _{PC}^{R}\rangle ,|\Phi _{PC}^{L}\rangle )=\mathcal{D}%
(|\Phi _{PC}^{+}\rangle ,|\Phi _{PC}^{-}\rangle )
\end{equation}%
These equations can be assumed as the theoretical conditions assuring the
same behaviour for any quantum MQS\ state generated by the QI-OPA\ in the
collinear configuration: they identify the equatorial set of the Bloch
sphere a privileged resilient to losses Hilbert subspace. The \textit{%
visibility} of the state $\left\vert \Phi _{PC}^{R/L}\right\rangle $ was
evaluated numerically analyzing the Bures distance as a function of $x$:\
Figure 11. Note that for small values of $x$ the decay of $\mathcal{D}(x)$
is far slower than for the coherent $\alpha -MQS$ case.

\begin{figure}[t!!]
\centering \includegraphics[width=0.4\textwidth]{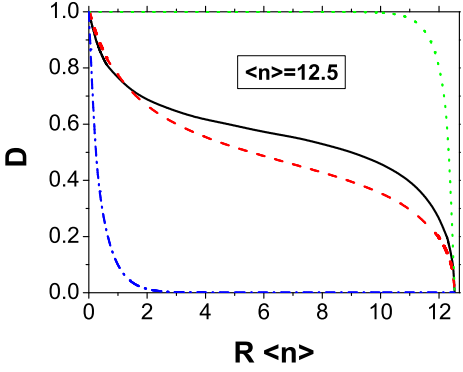}
\caption{Bures distance for various classes of MQSs for $\langle n\rangle
=12.5$. The lower blue dash-dotted curve corresponds to $\mathcal{D}\left(%
\mathcal{E}( |\protect\phi ^{+}\rangle) ,\mathcal{E}(|\protect\phi %
^{-}\rangle \right))$, while the green dotted upper curve is relative to the
distinguishability $\mathcal{D} (\mathcal{E}(|\protect\alpha^{+}\rangle), 
\mathcal{E}(|\protect\alpha^{-}\rangle)) $. The black straight curve
corresponds to the MQS generated by phase-covariant cloning $\mathcal{D}(%
\mathcal{E}(|\Phi _{PC}^{+}\rangle ),\mathcal{E}(|\Phi _{PC}^{-}\rangle ))$ 
\protect\cite{DeMa09,DeMa09a}, while the red dashed curve corresponds to the
universal cloning based MQS $\mathcal{D}(\mathcal{E}(|\Phi _{U}^{+}\rangle ),%
\mathcal{E}(|\Phi _{U}^{-}\rangle ))$ \protect\cite{spag10pra}.}
\label{fig:resilience}
\end{figure}

\subsection{Universal optimal quantum cloning machine}

Let's now investigate the resilience to decoherence of the MQS generated by
the universal optimal quantum cloning machine \cite{spag10pra}. At variance
with the phase-covariant amplifier, the output states do not exhibit any
comb structure in their photon number distributions. In agreement with the
universality property of the source, the Bures distance between the MQS
states $|\Phi _{U}^{1\psi }\rangle =\cos (\theta /2)|\Phi _{U}^{1H}\rangle
+e^{\imath \phi }\sin (\theta /2)|\Phi _{U}^{1V}\rangle $ and $|\Phi
_{U}^{1\psi _{\bot }}\rangle $ is independent of the choice of $(\theta
,\phi )$: 
\begin{equation}
\mathcal{D}(\hat{\rho}_{U}^{1\psi },\hat{\rho}_{U}^{1\psi _{\bot }})=%
\mathcal{D}(\hat{\rho}_{U}^{1\psi ^{\prime }},\hat{\rho}_{U}^{1\psi _{\bot
}^{\prime }})
\end{equation}%
for any basis $\left\{ \vec{\pi}_{\psi },\vec{\pi}_{\psi ^{\prime }}\right\} 
$. The larger symmetry of the latter identifies a larger Hilbert space of
macroscopic quantum superpositions resilient to decoherence, corresponding
to the complete polarization Bloch sphere. The cost of this larger symmetry
is a lower Bures distance in the universal case with respect to the
phase-covariant one. This represents an expected trade off in similar cases,
e.g. it parallels the well known increase of cloning fidelity due to the
reduced size of the Hilbert subspace in the case of phase covariance.

\subsection{Effective size of the multiparticle superposition}

Recent experiments on the formation of quantum superposition states in
near-macroscopic systems raise the question of how the sizes of general
quantum superposition states are to be quantified \cite{Legg02}. The first
method to quantify the Cat-size measure was introduced by \cite{Legg80}: the
so-called "`disconnectivity". However a closer analysis of the
disconnectivity shows that for indistinguishable particles this quantity is
large even for no-superposition states, like single-branch Fock states, due
to the particle correlations induced by symmetrization.

In the last few years, several criteria have been developped to establish
the effective size of macroscopic superpositions. \cite{dur02prl}
investigated state having the form $|\phi _{1}\rangle ^{\otimes M}+|\phi
_{2}\rangle ^{\otimes M}$ , where the number of subsystems $M$ is very
large, but the states of the individual subsystems have large overlap equal
to $1-\epsilon ^{2}$. These authors proposed two different methods for
assigning an effective particle number to such states, using ideal
Greenberger-Horne-Zeilinger states of the form $|0\rangle ^{\otimes
M}+|1\rangle ^{\otimes M}$ as a standard of comparison. The two methods,
based on decoherence and on a distillation protocol, lead to an effective
size $n$ of the order of $M\epsilon ^{2}$. The adoption of this criteria to
superconducting flux states provides a situation where counting the number
of electrons that are involved in the two current carrying the states gives
a very large estimate for the size of the superposition, while a detailed
analysis of how many electrons are actually behaving differently in the two
branches gives a very different and much smaller value. The Dur, Simon and Cirac criteria has been latter generalized by \cite{marq08pra}, which proposed a size measure based on counting how many single-particle operations are needed to map one state component (the "live cat") into the other one (the "died cat").  

A different approach has been introduced by \cite{kors07pra} who proposed a
measure of size for such superposition states that is based on what
measurements can be performed to probe and distinguish the different
branches of the macroscopic superposition. This approach allows the
comparison of the effective size for superposition states in very different
physical systems. Comparison with measure based on analysis of coherence
between branches \cite{Legg80} indicates that this measurement-based measure
provide significantly smaller effective superposition sizes. This criteria
have been applied to macroscopic superposition states in flux qubits
revealing the effective size to be bounded by values in the range of \ $%
12\div 100$ \cite{kors10epl}.

While \cite{dur02prl} approach could not be applied to the present
amplification scheme, \cite{kors07pra} criteria for effective size can be
estimated by exploiting the previous results on the Bures distance between
the macro-states. The problem of determining quantum states that can be
deterministically discriminated can be directly related to the Bures
distance between the involved state \cite{mark08pra}. There it has been
shown that the probability of success of the discrimination $p_{disc}$
between two states ${\rho _{1},\rho _{2}}$ fullfill the bound $p_{disc}(\rho
_{1},\rho _{2})\leq \mathcal{D}(\rho _{1},\rho _{2})$. According to Figure
11, the Bures distance $\mathcal{D}(\mathcal{E}(|\Phi _{PC}^{+}\rangle ),%
\mathcal{E}(|\Phi _{PC}^{-}\rangle ))$ between falls down from $1$ to $0.8$
as soon as an average of 1 photon is lost. Accordingly the close to perfect
discrimination among the states $|\Phi _{PC}^{+}\rangle $ and $|\Phi
_{PC}^{-}\rangle $ would require to detect almost all particles. This result
suggests that according to \cite{kors07pra} criteria the effective size of
the macroscopic quantum superposition is rather limited, analogously to
superconducting macro-superposition. On the other side, we should note that
the macrostates, $\left\vert \Phi ^{\phi }\right\rangle _{B}$, $\left\vert
\Phi ^{\phi \perp }\right\rangle _{B}$ exhibit observables bearing
macroscopically distinct average values even in lossy regime with
transmittivity $T$. Precisely, the average number of photons associated with
the polarization mode $\overrightarrow{\pi }_{\phi }$ is: $T\overline{m}$
for $\left\vert \Phi ^{\phi \perp }\right\rangle _{B}$, and $T(3\overline{m}%
+1)$ for $\left\vert \Phi ^{\phi }\right\rangle _{B}$. For the $\pi -$mode $%
\overrightarrow{\pi }_{\phi \perp }$ , orthogonal to $\overrightarrow{\pi }%
_{\phi }$, these values are interchanged among the two states. Hence we tend
to agree with\cite{kors07pra} according to which, that more general measures
for comparing the effective size of superposition states in different kinds
of physical systems should be developed.

Other approaches have been proposed to quantify macroscopic quantum superposition. First of all \cite{Bjo04} proposed an operational approach. Their size criterion for macroscopic superposition states is based on the fact that a superposition presents greater sensitivity in interferometric applications than its superposed constituent states.  \cite{Lee11} proposed to quantify the degree of quantum coherence and the effective size of the physical system that involves the superposition by exploiting quantum interference in phase space. Finally  \cite{shim02prl,shim05prl} proposed an index of macroscopic entanglement based on correlation of local observables on many sites in macroscopic quantum systems.

\section{WIGNER - FUNCTION THEORY}

Let us now address the problem of providing a complete quantum phase-space
analysis able to recognize the persistence of the QI-OPA \ properties in a
decohering environment. Among the different representation of quantum states
in the continuous-variables space \cite{Cahi69}, the Wigner
quasi-probability representation has been widely exploited to investigate
non-classical properties, such as squeezing \cite{Wall95} and EPR
non-locality \cite{Bana98}. In particular, the presence of negative
quasi-probability regions has been considered as a consequence of the
quantum superposition of distinct physical states \cite{Bart44}. By the way,
the negativity of the Wigner function is not the only parameter that allows
to estimate the non-classicality of a certain state. For instance, the
squeezed vacuum state \cite{Wall95} presents a positive $W$-representation,
while its properties cannot be described by the laws of classical physics.
Furthermore, recent papers have shown that the Wigner function of an EPR
state provides direct evidence of its non-local character \cite%
{Cohe97,Bana98}, while being completely-positive in all the phase-space.

In order to investigate the properties of the output field of the QI-OPA
device in details, we analyze the quasi-probability distribution introduced
by Wigner \cite{Wign32} for the amplified field. The Wigner function is
defined as the Fourier transform of the symmetrically-ordered characteristic
function $\chi (\eta )$ of the state described by the general density matrix 
$\hat{\rho}$ 
\begin{equation}
\chi (\eta )=Tr\left[ \hat{\rho}\exp \left( \eta \hat{a}^{\dagger }-\eta
^{\ast }\hat{a}\right) \right]  \label{funz_caratt}
\end{equation}%
The associated Wigner function 
\begin{equation}
W(\alpha )=\frac{1}{\pi ^{2}}\int \exp \left( \eta ^{\ast }\alpha -\eta
\alpha ^{\ast }\right) \chi \left( \eta \right) d^{2}\eta  \label{deffunzwig}
\end{equation}%
exists for any $\hat{\rho}$ but is not always positive definite and,
consequently, can not be considered as a genuine probability distribution.

\begin{figure}[tbp]
\centering \includegraphics[width=0.5\textwidth]{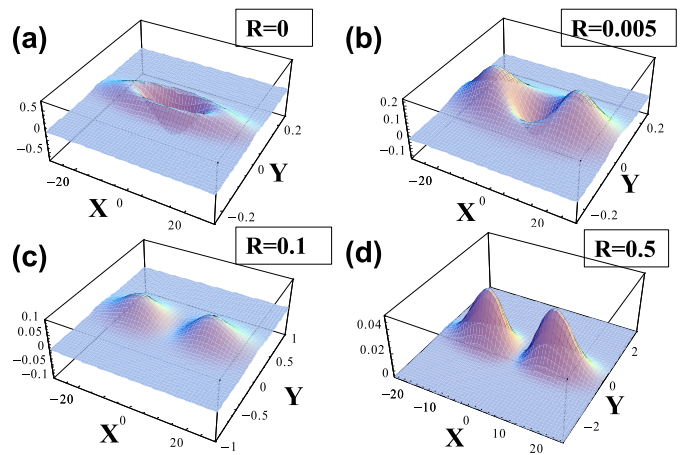}
\caption{Wigner function of a single-photon amplified state in a single-mode
degenerate OPA for $g=3$. (a) (R=0) Unperturbed case. (b) (R=0.005) For
small reflectivity, the Wigner function remains negative in the central
region. (c) (R=0.1) The Wigner function progressively evolve in a positive
function in all the phase-space. (d) (R=0.5) Transition from a non-positive
to a completely positive Wigner function \protect\cite{Spag09}.}
\label{fig:single_mode_1_photon_amplified_losses}
\end{figure}

The properties of the multiphoton system have been investigated \cite{Spag09}
in phase-space by a Wigner quasi-probability function analysis when the
fields propagate over a lossy channel. Fig.\ref%
{fig:single_mode_1_photon_amplified_losses}-a reports the ideal case, in
absence of losses, showing the presence of peculiar quantum properties such
as squeezing and a non-positive W-representation. Then, by investigating the
resilience to losses of QI-OPA amplified states in a lossy configuration,
the persistence of the non-positivity of the Wigner function was
demonstrated in a certain range of the \emph{system-enviroment} interaction
parameter $R$. This behaviour can be compared with the one shown by the $%
|\alpha \rangle $ states MQS, which features a non-positive W-representation
in the same interval of $R$. The more resilient structure of the QI-OPA
amplified states has been enlightened by their slower decoherence rate,
represented by both the slower decrease in the negative part of the Wigner
function and by the behavior of the Bures distance between orthogonal
macrostates \cite{Spag09}. Since the negativity of the W-representation is a
sufficient but not a necessary condition for the non-classicality of any
physical system, future investigations should be aimed at the regime of
decoherence in cases in which the Wigner function is completely positive,
analyzing by different criteria the presence of the related quantum
properties of the system.

\section{GENERATION OF MACRO-MACRO ENTANGLED STATES}

One of the main challenges for an experimental test of entanglement in
systems of large size is the realization of suitable criteria for the
detection of entanglement in bipartite macroscopic systems. A large effort
has been devoted in the last few years in this direction \cite{horo09rmp}.
Some criteria, such as the partial transpose criterion developed by \cite%
{pere96prl,horo96pla}, require the tomographic reconstruction of the density
matrix, which from an experimental viewpoint is generally highly demanding
for system composed by a large number $M$ of particles. However, the
complete reconstruction of the state can be avoided by the \textquotedblleft
entanglement witness\textquotedblright\ method consisting of a class of
tests where only few significant local measurements are performed. For
bipartite systems with large $M$, this approach has been applied via
collective measurements on the state. Within this context, Duan et al.
proposed a general criterion based on measurements on "continuous variables"
observables\cite{Duan00,Brau05}. This general criterion was subsequently
applied to the quantum extension of the Stokes parameters in order to obtain
an entanglement bound for such type of variables \cite{Koro02,Koro05,Schn03}%
. Other approaches have been developed based on spin variables \cite{Simo03}
or pseudo-Pauli operators \cite{Chen02}. An experimental application of this
criteria based on collective spin measurements has been performed in a
bipartite system consisting by separate two gas samples \cite{Juls01}.


The main experimental problem for such observations arises from the
requirement of attaining a sufficient isolation of the quantum system from
its environment, i.e., from the decoherence process \cite{Zure03}. An
alternative approach to explain the quantum-to-classical transition has been
recently proposed by Kofler and Brukner, along an idea earlier discussed by
Bell, Peres and Mermin \cite{Pere93}. They\ considered the emergence of \
classical physics in systems of increasing size \textit{within }the domain
of quantum theory \cite{Kofl07}. Precisely, they focused on the limits of
the observability of quantum effects in macroscopic objects, showing that,
for large systems, macrorealism arises under coarse-grained measurements.
However, some counterexamples to such modellization were found later by the
same authors: some non classical Hamiltonians violate macrorealism in spite
coarse-grained measurements \cite{Kofl08}. Therefore the problem of the
resolution within the measurement process appears to be a key ingredient in
the understanding the limits of the quantum behavior of macroscopic physical
systems and the quantum-to-classical transition. In a recent paper Jeong et
al.contributed to the investigation about the possibility of observing the
quantum features of a system under fuzzy measurement, by finding that
extremely-coarse-grained measurements can still be useful to reveal the
quantum world where local realism fails\cite{Jeon09}.


\begin{figure}[h]
\centering \includegraphics[scale=0.28]{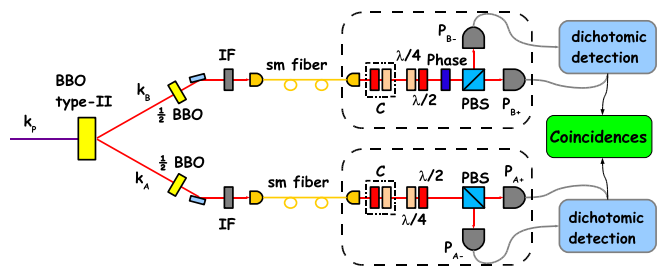}
\caption{Setup for the generation and detection of a bipartite macroscopic
field. The high laser pulse on mode $\mathbf{k}_{P}$ excites a type-II
noncolinear source in the high gain regime ($g=3.5$). The two spatial mode $%
\mathbf{k}_{A}$ and $\mathbf{k}_{B}$ are spectrally and spatially selected
by interference filters (IF) and single mode fibers. After fiber
compensation $(C)$, the two modes are analyzed in polarization and detected
by four photomultipliers \protect\cite{Vite10}. }
\label{fig:experimental_setup_spont}
\end{figure}

\subsection{Macroscopic quantum state based on high gain spontaneous
parametric down-conversion}

\label{sec:Macro_Macro_system} Let us consider, once again, an optical
parametric amplifier working in a high gain regime. The radiation field
under investigation is the quantum state obtained by spontaneous parametric
down-conversion (SPDC) \cite{Kwia95,Eise04}, whose interaction Hamiltonian
is: $\mathcal{H}_{U}=\imath \hbar \chi \left( \hat{a}_{\pi }^{\dag }\hat{b}%
_{\pi _{\bot }}^{\dag }-\hat{a}_{\pi _{\bot }}^{\dag }\hat{b}_{\pi }^{\dag
}\right) +\mathrm{H.c.}$.

The output state reads \cite{Simo03,Eise04,Cami06}: 
\begin{equation}
|\Psi ^{-}\rangle =\frac{1}{C^{2}}\sum_{n=0}^{\infty }\Gamma ^{n}\sqrt{n+1}%
||\psi _{n}^{-}\rangle  \label{eq:SPDC_state}
\end{equation}%
with {\small 
\begin{equation}
|\psi _{n}^{-}\rangle =\frac{1}{\sqrt{n+1}}\sum_{m=0}^{n}(-1)^{m}|(n-m)\pi
,m\pi _{\bot }\rangle _{A}|m\pi ,(n-m)\pi _{\bot }\rangle _{B}
\label{eq:singlet_n}
\end{equation}%
}The output state can be written as the weighted coherent superposition of
singlet spin-$\frac{n}{2}$ states $|\psi _{n}^{-}\rangle $.

This source has been adopted in many experiments, at different gain regimes.
First, \cite{Kwia95} exploited the polarization singlet-state emitted in the
single-pair regime to test the violation of Bell's inequalities. Further
work demonstrated experimentally the four-photon entanglement in the
second-order emission state of the SPDC source, by detecting the four-fold
coincidences after the two output modes of the source were coupled to two
50-50 beam-splitters (BS)\cite{Eibl03}. Moreover, a generalized non-locality
test was also successfully performed with this configuration\cite{Wein01}.
Later, a similar scheme was adopted by Wieczorek et al. to experimentally
generate an entire family of four-photon entangled states\cite{Wiec08}.

\subsubsection{Non-separable Werner states}

As previously mentioned, the presence of polarization-entanglement in the
multi-photon states up to $M=12$ photons was experimentally proved by
investigating the high loss regime in which at most one photon per branch
was detected \cite{Eise04,cami06pra}. This approach consisted of the
generation of a multiphoton state followed by a strong attenuation on both
output branches of the SPDC scheme, in order to \textquotedblleft extract'\
a correlated couple of photons, one for each branch:\ Figure 13. The method
presents several advantages: first, the techniques for single-photon
detection and characterization can be adopted. Second, it models the effect
of loss associated with any communication process on a multiphoton entangled
state.

The density matrix of the two-photon state has been investigated by theory
and experiment\cite{Cami06}. The state given by Eq.(\ref{eq:SPDC_state}), is
stochastically attenuated by a conventional beam-splitter model that
simulates the propagation over a lossy channel. Then the density matrix of
the two-photon state generated by postselection is expressed by: 
\begin{equation}
\rho _{SPDC}^{HG}=\left( 
\begin{array}{cccc}
\frac{1-p}{4} & 0 & 0 & 0 \\ 
0 & \frac{1+p}{4} & -\frac{p}{2} & 0 \\ 
0 & -\frac{p}{2} & \frac{1+p}{4} & 0 \\ 
0 & 0 & 0 & \frac{1-p}{4}%
\end{array}%
\right)  \label{Wernerstate}
\end{equation}%
with singlet weight $p=\frac{1}{2\widetilde{\Gamma }^{2}+1}\label{Weight}$
and $\widetilde{\Gamma }=(1-\eta )\tanh g$. We note that the density matrix $%
\rho _{SPDC}^{HG}$ is a Werner state, i.e., a weighted superposition of a
maximally entangled singlet state with a fully mixed state \cite{wern89pra}.
As it is well known, the Werner states play a paradigmatic role in quantum
information; as they determine a family of mixed states including both
entangled and separable states \cite{barb04prl}. They model the decoherence
process occurring on a singlet state traveling along a noisy channel, and
hence they are adopted to investigate the distillation and concentration
processes. Furthermore, depending on the singlet weight they can exhibit
either entanglement and violation of Bell inequalities, or only
entanglement, or separability. In the limit $\eta \rightarrow 0$ the above
equation gives: $\widetilde{\Gamma }=\tanh g$ $\approx 1$, for large $g$. In
the hypothesis of very high losses, the singlet weight $p\geq \frac{1}{3}$
approaches the minimum value $\frac{1}{3}$. Since the condition $p>\frac{1}{3%
}$ implies the non-separability condition for a general Werner state, the
two-photon state is entangled for any large value of g. Figure 13 shows the
result of the theory together with the experimental demonstration of
bipartite entanglement for: $M\leq 12$.

\begin{figure}[h]
{\small \centering 
\includegraphics[width=0.35\textwidth,
angle=270]{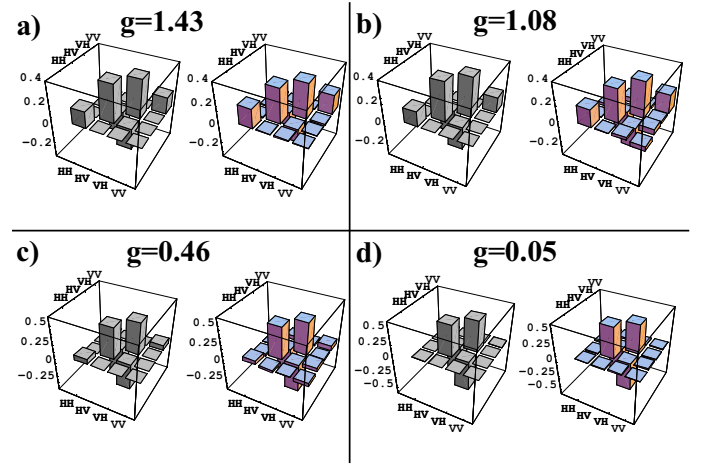} }
\caption{Theoretical (left plot) and experimental(right plot) density
matrices $\protect\rho _{SPDC}^{HG}$ for different gain values. The
experimental density matrices have been reconstructed by measuring $16$ two
qubits observables \protect\cite{cami06pra}.}
\label{fig:experimental_setup_spont}
\end{figure}

\subsubsection{Quantum-to-classical transition by dichotomic measurement}

We are now interested in analyzing the behavior of the system above
considered when the number of generated photons is increased and the system
undergoes a fuzzy dochotomic measurement on the overall state, in which the
generated particles cannot be addressed singularly. As shown by \cite{Chen02}%
, the demonstration of non-locality in a multiphoton state produced by a
non-degenerate optical parametric amplifier would require the experimental
application of "parity operators", with a detector efficiency: $\eta =1$. On
the other hand, the estimation of a coarse grained quantity through
collective measurements as suggested by \cite{Port06}, would miss the
underlying quantum structure of the generated state, introducing elements of
local realism even in presence of strong entanglement and in absence of
decoherence. A theoretical investigation on a multiphoton system generated
by parametric down conversion was carried out by Reid and coworkers \cite%
{Reid02}. They analyzed the possibility of testing the violation of Bell's
inequality by performing dichotomic measurement on the multiparticle quantum
state. Precisely, in analogy with the spin formalism and the O-Filter
discrimination, they proposed to compare the number of photons polarized
\textquotedblleft up\textquotedblright\ with the number of photons polarized
\textquotedblleft down\textquotedblright\ at the exit of the amplifier: a
dichotomic measurement on the multiphoton state. In such a way a small
violation of the multiparticle Bell's inequality can be revealed even in
presence of losses and of the quantum inefficiency of detectors. Once again,
the violation decreases very rapidly for an increasing number $M$ of the
generated photons. In a recent paper \cite{Banc08} have discussed different
techniques for testing the Bell's inequality violation in multipair
scenarios by performing a global measurement, in either Alice's and Bob's
sites. According to their theory, the photon pairs were classified as
"distinguishable", i.e. independent, or "indistinguishable", i.e. belonging
to the same spatial and temporal mode. They found that while the state of
indistinguishable pairs results more entangled, the state of independent
pairs appears to be more nonlocal.

The possibility of observing quantum correlations in macroscopic systems
through dichotomic measurement, by addressing two different measurement
schemes, based on different dichotomization processes has recently addressed
by \cite{Vite10}. More specifically, the persistence of non-locality in a
spin-$\frac{n}{2}$ singlet state with increasing size has been investigated
by studying the change in the correlations form as $n$ increases, both in
the ideal case and in presence of losses. Two different types of dichotomic
measurements on multiphoton states were considered: the orthogonality
filtering and the threshold detection. Numerical simulation showed that the
interference fringe-patterns for singlet-$\frac{n}{2}$ states exhibit a
transition from the sinusoidal pattern of the spin-$\frac{1}{2}$ into a
quasi-linear pattern by increasing the number of photons associated with the
spin state. According to this behavior a progressive decrease of the amount
of the violation is observed, as earlier predicted by \cite{Reid02,Banc08}.
All these results show that the dichotomic fuzzy measurements lack of the
necessary resolution to characterize such states. They also show, once
again, how problematic is the experimental demonstration of quantum
non-locality of states with very large $M$.

\subsection{Macroscopic quantum state by dual amplification of two-photon
entangled state}

\begin{figure}[t]
{\small \includegraphics[scale=.4]{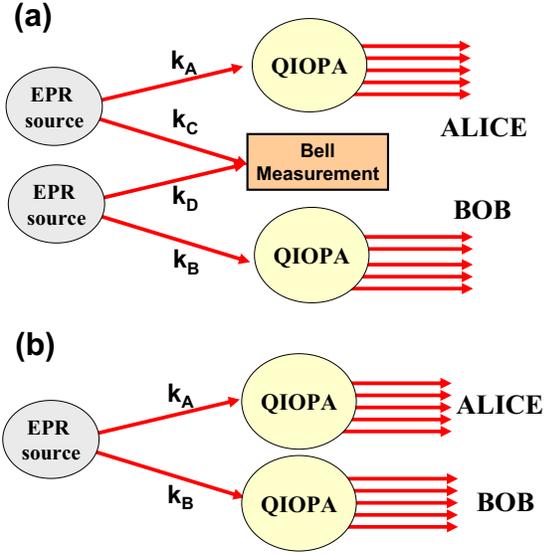} }
\caption{Implementation of macro-macro entanglement (a)\ via entanglement
swapping on two QI-OPA (b) through two optical parametric amplifiers 
\protect\cite{DeMa10fp}.}
\end{figure}

The amplification schemes illustrated by the Figures 1 a), b), c) could be
upgraded in order to achieve an entangled macro-macro system showing
nonlocality features \cite{DeMa10fp}. Such scheme could\ even exploit an
entanglement swapping protocol as shown in Fig.8 a)\cite{zuko93prl,pan98prl}%
. There the final entangled state is achieved through a standard
intermediate Bell measurement carried out on the single photon states. A
similar process has been suggested in several different contexts, e.g. to
entangle micromechanical oscillators \cite{Pira06}. As an alternative
approach, the single photon states on mode $k_{A}$ and $k_{B}$ could be
amplified by two independent QI-OPA's :Fig.8 b). The resulting Macro-Macro
scenario would be an interesting platform to perform loophole free Bell
inequalities.

An open question is how to perform an entanglement or/and non-locality test
on the Macro-Macro states. Indeed, analogously to the Micro-Macro scenario,
a coarse-grained measurement resolution would be requested. To overcome this
challenge, it has been proposed to manipulate multiphoton quantum states
obtained through optical parametric amplification by performing a
measurement on a small portion of the output light field. \cite{vite10pra}
analyzed in details how the quantum features of the Macro-states are
modified by varying the amount of extracted information and considered the
best strategy to be adopted at the final measurement stage. At last it was
found that the scheme does not allow one to violate any multiphoton Bell's
inequality in absence of auxiliary assumptions.

A similar investigation on the preprocessing of quantum macroscopic states
of light generated by optimal quantum cloners in the presence of classical
detection has been carried out by \cite{stob11pra}. These Authors proposed a
filter that selects two-mode high number Fock states whose photon-number
difference exceeds a certain value. This filter improves the
distinguishability of some states by preserving the quantum macroscopic
superposition \cite{stob11c}. It is still an open question whether this
filter can be efficiently implemented and whether it can lead to a genuine
non-locality test.

\section{INTERACTION WITH A BOSE-EINSTEIN CONDENSATE}

\begin{figure}[t]
{\small \includegraphics[width=8 cm]{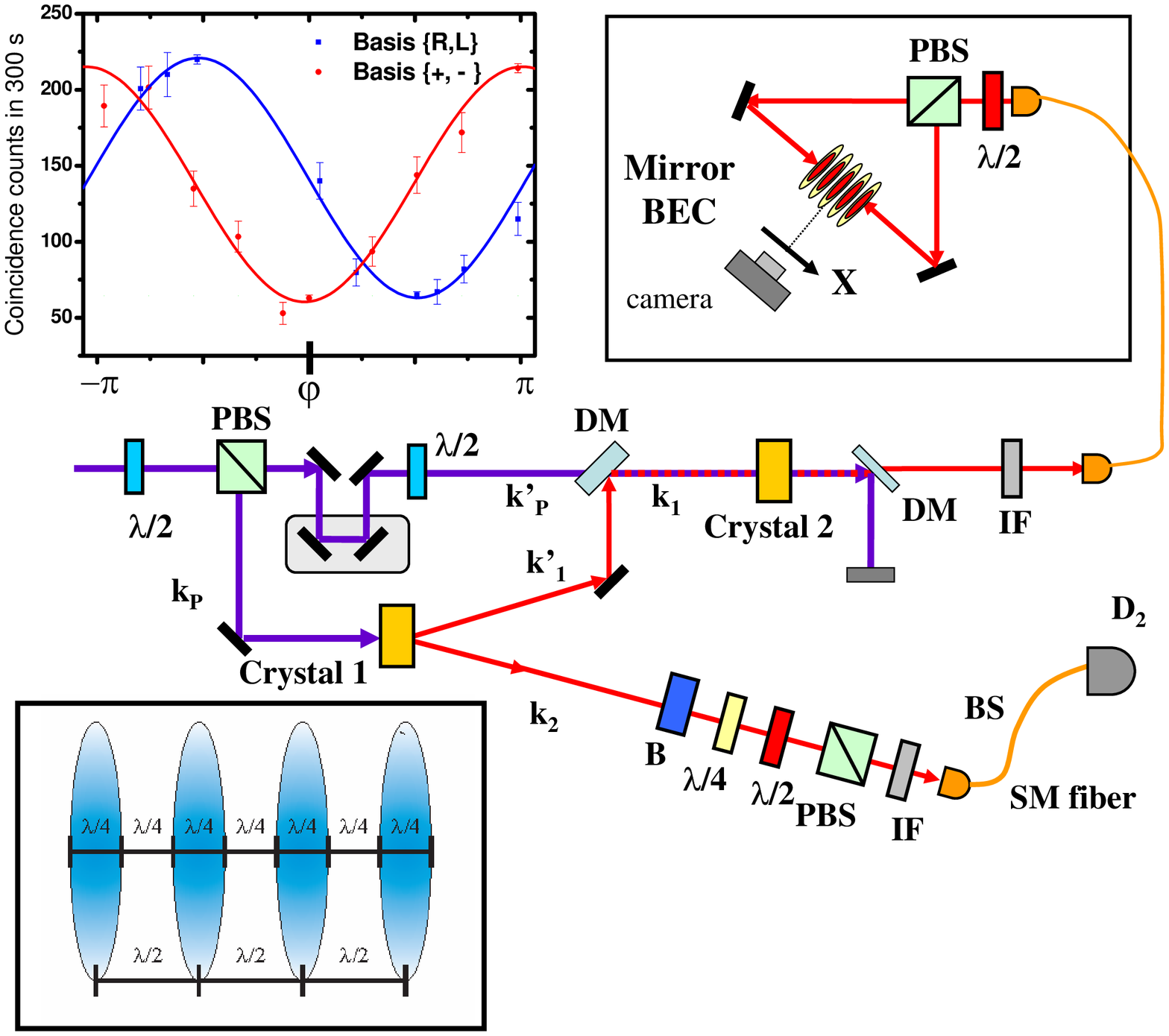} }
\caption{Layout of the QI-OPA and Mirror BEC experimental apparatus. The
upper left inset shows the interference\ patterns detected at the output of
the PBS\ shown in the\ UR-Inset for two different measurement basis $\left\{
+,-\right\} $ and $\left\{ L,R\right\} $. Alternating slabs of condensate
and vacuum are shown in the lower left inset \protect\cite{DeMa10}}
\label{schema}
\end{figure}

In recent years a great deal of interest has been attracted by the ambitious
challenge of creating a macroscopic quantum superposition of a massive
object by an entangled opto-mechanical interaction of a tiny mirror with a
single photon trapped within a Michelson interferometer \cite{mars03prl},
This would lead to another realization of the well known 1935 argument by
Schr\"{o}dinger \cite{Schr35}. A similar scheme could be considered which is
based on the nonresonant scattering by a properly shaped \textit{multi-atom }%
Bose--Einstein condensate (BEC) of the \textit{multi-photon} state $%
\left\vert \Phi \right\rangle $ generated by a high-gain quantum-injected
optical parametric amplifier (QI-OPA) described at Chapter II\ of the
present article. Light scattering from BEC structures has been adopted so
far to enhance their non-linear macroscopic properties in super-radiance
experiments \cite{Inouye99}, to show the possibility of matter wave
amplification \cite{Kozuma99} and non-linear wave mixing \cite{Deng99}. The
new scheme, represented by Fig.\ref{schema}, would result in a joint
atom-photon Micro-Macro state entangled by momentum conservation. The
resulting physical effect would consist of the mechanical motion of a high
reflectivity optical multilayered Bragg-shaped mirror, referred to as a
\textquotedblright Mirror-BEC\textquotedblright , driven by the exchange of
linear momentum with a photonic Macrostate $\left\vert \Phi \right\rangle $.

The layout in Figure \ref{schema} shows a QI-OPA system identical to the one
represented in Figure 1 c). The interfereing polarization macrostates
belonging to the quantum superposition (MQS): $\left\vert \Phi \right\rangle
=2^{-%
{\frac12}%
}(\left\vert \Phi ^{\phi }\right\rangle +\left\vert \Phi ^{\phi \perp
}\right\rangle )$ generated over mode $k_{2}$ are selected by a polarizing
beam splitter and drive along the X-axis\ the mechanical motion of the
Mirror-BEC. Precisely, the displacements along the two opposite directions
parallel to the X-axis are driven respectively by the orthogonal
polarizations:$\ \left\vert \Phi ^{\phi }\right\rangle $ and $\left\vert
\Phi ^{\phi \perp }\right\rangle $. \ Since these states are found to be
entangled with the far apart single-photon emitted over the mode $k_{2}$,
the same entanglement property can be transferred to the position-Macrostate
of the optically-driven Mirror-BEC. The discussion in Section V dealing with
the entanglement processes can be extended to the present more complex
opto-mechanical configuration.

\section{APPLICATIONS: FROM SENSING TO RADIOMETRY}

\subsection{Quantum sensing}

The aim of quantum sensing is to develop stategies able to extract from a
system the maximum amount of information with a minimal disturbance. The
possibility of performing precision measurements by adopting quantum
resources can increase the achievable precision going beyond the
semiclassical regime of operation \cite{Giov04b,Giov06,Hels76}. In the case
of interferometry, this can be achieved by the use of the so-called N00N
states, which are quantum mechanical superpositions of just two terms,
corresponding to all the available photons $N$ placed in either the signal
arm or the reference arm of the interferometer. The use of N00N states can
enhance the precision in phase estimation to $1/N$, thus improving the
scaling of the achievable precision with respect to the employed resources 
\cite{Boto00,Dowl08}. This approach can have wide applications for minimally
invasive sensing methods acting on quantum states. Nevertheless, these
states result extremely fragile under unavoidable losses and decoherence\cite%
{Gilb08}. For instance, a sample, whose phase shift is to be measured,
generally introduce attenuation. Since the quantum-enhanced modes of
operations are generally very fragile the impact of environmental effects
can be much more harmful than in semiclassical schemes by destroying
completely the quantum benefits \cite{Rubi07,Shaj07}. This scenario explains
why the overcoming the negative effects of \ the realistic environments is
the main challenge of the technology of quantum sensing. Very recently, the
theoretical and experimental engineering of quantum states of light has
attracted much attention, leading to the best possible precision in optical
two-mode interferometry, even in presence of experimental imperfections \cite%
{Huve08,Dorn09,Macc09,Demk09,Kacp09,Lee09}.\newline
\begin{figure}[t]
{\small \includegraphics[width=0.49\textwidth]{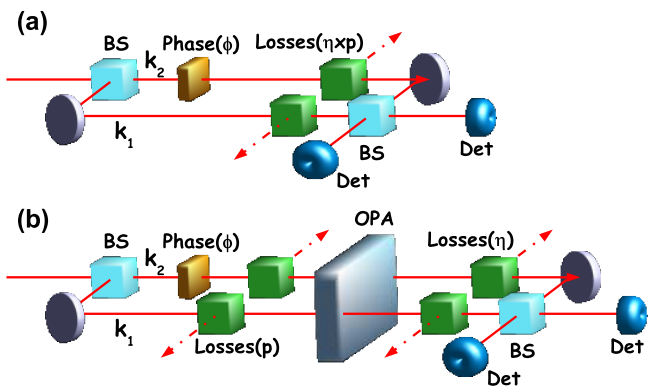} }
\caption{Scheme for the phase measurement. (a) Interferometric scheme
adopted to estimate the phase ($\protect\phi $) introduced in the mode $%
\mathbf{k}_{2}$. (b) Interferometric scheme adopting a single photon and the
optical parametric amplifier: the amplification of the single photon state
is performed before dominant losses \protect\cite{vite10prl}.}
\label{fig:schema_concettuale}
\end{figure}

Recently \cite{vite10prl} reported an hybrid approach based on a high gain
optical parametric amplifier operating for any polarization state in order
to transfer quantum properties of different microscopic quantum states in
the macroscopic regime. By performing the amplification process of the
microscopic probe after the interaction with the sample, it is possibile to
overcome the detrimental effects of losses on the phase measurement which
affects the single photon state following the test on the sample. This
approach may be adopted in a minimally invasive scenario where a fragile
sample, such as biological system requires a minimum amount of test photons
in order to prevent damages. The action of the amplifier, i.e. the process
of optimal phase covariant quantum cloning, is to amplify the phase
information which is codified in a single photon into a large number of
particles. Such multiphoton states exhibit a high resilience to losses, as
shown by \cite{DeMa08,DeMa09,DeMa09a}, and can be manipulated by exploiting
a detection scheme which combines features of discrete and continuous
variables. The effect of losses on the macroscopic field consists of the
reduction of the detected signal and not in the complete cancellation of the
phase information as in the single photon probe case, thus improving the
achievable sensitivity. This improvement consists in a constant enhancement\ 
$K(g)$ of the sensitivity, depending on the amplifier gain $g$. Hence, the
sensitivity scales as $\sqrt{N}$, where $N$ is the number of photons testing
the sample, but the effect of the amplification process reduces the
detrimental effect of losses by a factor proportional to the number of
generated photons. Within this frameworkrecently a \cite{Eschl11} derived the general bounds on the adoption of quantum metrology in the presence of decoherence are obtained.

\subsection{Quantum radiometry}

Radiometry is the science of measuring the electromagnetic radiation. The
available techonologies in this field can operate either in the relative
high power regime or in the photon-counting regime based on the correlations
of quantum fields. Very recently a sophisticated radiometer apparatus has
been devised that works over a broad range of powers: from the single photon
level, up to several tens of nW, i.e. from the quantum regime to the
classical regime \cite{sang10prl}. In fact, such system exploits the process
of optimal quantum cloning and is able to provide an absolute measure of
spectral radiance by relying on a particular aspect of the
quantum-to-classical transition: as the number of information carriers
(photons) grows, so does the cloning fidelity. Sanguinetti et al. have shown
that the fidelity of cloning can be used to produce an absolute power
measurement with an uncertainty limited only by the uncertainty of a
relative power measurement. They provided a convincing demonstration of this
scheme by an all-fiber experiment at telecom wavelengths, by achieving an
accuracy of 4\%, a figure that can be easily improved by a dedicated
metrology laboratory\cite{fase02prl}.

\section{CONCLUSIONS AND PERSPECTIVES}

\bigskip In this paper we have reviewed several protocols and related
experiments centered on the process of nonlinear amplification of single
photon quantum states. A large part of the investigation focused on the new
protocol of quantum information, viz. the "quantum injected" version of the
optical parametric amplification (OPA), by which a single photon, encoded as
a Micro-qubit, "triggers" by a QED\ process the generation of an in
principle unlimited number $M$ of photons, \ i.e a Macro-qubit, carrying a
large portion of the information associated with the trigger particle. The
highly seminal character of the new protocol opened the way to the
discovery, the realization and the developement of novel scientific methods
and applications of \ fundamental and technical relevance. The quantum
information protocols today generally referred to as "quantum \ cloning",
"quantum U-NOT", "macroscopic quantum superposition" (MQS), "Micro-Macro
entanglement","quantum reversion", "quantum-to-classical transition" were
amongst the paradigmatic outcomes of the overall endeavour reported in this
article, lasting more than one decade. In particular, the QI-OPA method was
instrumental for the first experimental realization of the\
"quantum-cloning" process in several multiparticle regimes. This process was
further thoroughly investigated leading to the discovery of the U-NOT\
theorem, of the quantum reversion protocol and to the first experimental
test of the "no signalling theorem". In this connection, the unexpected
result of the experiment was that the impossibility of "faster than light
communication", i.e. the according to Abner Shimony "peaceful coexistence"
between Special Relativity and quantum mechanics, rests of the high-order
correlations affecting the particles generated by a cloning machine.

A large part of the investigation focused on the realization via quantum
cloning of the Macroscopic Quantum Superposition (MQS) process, which is
related to the outstanding quantum-to-classical transition paradigm implied
by the celebrated "\textit{Schr\H{o}dinger's Cat}" argument\cite{Schr35}.
The realization of the MQS\ process consisting of a large number of
particles $M\geq 10^{3}\ $was experimentally demonstrated, in a \textit{%
non-entangled} configuration, by detection of the sinusoidal phase
dependence of the interference fringe patterns generated at the output of
the apparatus. However the bipartite Micro-Macro entanglement could be
demonstrated only for a reduced number of particles, $M\preceq 12$ owing to
the existence of a "detection loophole" whose detrimental effect increases
with $M$. A most interesting feature the adopted MQS\ scheme was found to
consist of its resilience to any externally driven de-coherence process:
this allowed to carry out the entire research at $T=300%
{{}^\circ}%
K$. Accordingly, a large emphasis has been given to an extended theoretical
analysis aimed at the understanding this phase-impairing process affecting
all multiparticle systems. \ The extension of the quantum-cloning argument
to a novel Macro-Macro regime and the mechanical coherent interaction of a
multi-photon MQS\ system with a multi-atom BEC\ condensate were considered
as proposals towards further research on the foundations of Quantum
Mechanics. 

Concerning the investigation on quantum-to-classical transition a
set of entanglement criteria for bipartite systems of a large number of
particles were introduced and analyzed in details. In particular, a specific
joint Micro- and Macro-scopic system based on optical parametric
amplification of an entangled photon pair was addressed. Their potential
applications of fundamental and technical relevance in different contexts
were analyzed, e.g. the realization of non-locality tests, quantum
metrology, quantum sensing and, as an open challenge for future research,
the process of "pre-selection", i.e. the establishment of efficient
strategies\ able to generate and wisely manipulate multiphoton states by
performing measurements on a small portion of the output field\cite%
{vite10pra}. Precisely, within the Macro-Macro nonlocality test, the aim was
to understand how the features of the Macro-qubit in the high-loss and
large-photon-number regime are modified by varying the amount of extracted
information and then to devise the best strategy to be adopted at the final
measurement stage. In facts, the proposed pre-selection method, the simplest
one based on the\ dichotomic measurement of the reflected part of the
wave-function in two different bases did not allow to violate a Bell's
inequality. At last a more general approach to the Micro-Macro entanglement
problem, based on single photon-continuous variables hybrid methods, was
introduced\cite{Spag11}. All these novel criteria and methods were
considered and compared in the context of the existing literature in the
feld.

In summary, we do believe that the extended theoretical and experimental
investigation outlined by the present article can contribute to open new paths of research either by stimulating the discovery of
efficient theorems and protocols of quantum information, and, on the more
fundamental side, by shedding new light on the still uncertain border
existing between the "classical" and the "quantum" aspects of Nature.

\section*{Acknowledgments}

We would like to thank Nicol\`o Spagnolo, Chiara Vitelli, Nicolas Gisin, Christoph Simon, Pawel Horodecki for interesting and enlightening discussions.  This work was supported by FIRB-Futuro in Ricerca (HYTEQ).


{\small \newpage }

\end{document}